\documentclass[preprint2,trackchanges, twocolumn]{aastex701}

\usepackage{mathtools} 
\usepackage{xspace}
\usepackage{array}

\def\code#1{\texttt{#1}}
\usepackage{tabularx}
\usepackage{array} 
\newcolumntype{C}{>{\centering\arraybackslash}X}
\newcommand{\mbh}{\ensuremath{{M_{\rm BH}}}\xspace}
\newcommand{\mstar}{\ensuremath{{M_\star}}\xspace}
\newcommand{\msun}{\ensuremath{{M_\odot}}\xspace}
\newcommand{\ocen}{\ensuremath{\omega {\rm Cen}}\xspace}

\newcommand{\kms}{\mbox{$\>{\rm km\, s^{-1}}$}}

\begin{document}

\newcommand{\adast}{AdAst}
\newcommand{\joss}{JOSS}
\newcommand{\jkas}{JKAS}
\newcommand{\natas}{NatAs}
\newcommand{\natme}{NatMe}

\title{\texttt{GASTRO-BH}: Effects of a Central Black Hole in GSE-like Dwarf-galaxy Mergers \\ Around Milky Way-like Hosts}

\author[orcid=0000-0002-9318-9329]{Pedro Naethe Motta}
\affiliation{Instituto de Astronomia, Geof\'{\i}sica e Ci\^encias Atmosf\'ericas, Universidade de S\~ao Paulo, S\~ao Paulo, SP 05508-090, Brazil.}
\affiliation{Black Hole Initiative at Harvard University, 20 Garden Street, Cambridge, MA 02138, USA}
\email[show]{pedronaethemotta@usp.br}

\author[orcid=0000-0002-7662-5475]{Jo\~{a}o A. S. Amarante}
\affiliation{Department of Astronomy, School of Physics and Astronomy, Shanghai Jiao Tong University, 800 Dongchuan Road, Shanghai, 200240, China}
\affiliation{State Key Laboratory of Dark Matter Physics, School of Physics and Astronomy, Shanghai Jiao Tong University, Shanghai, 200240, China}
\email[show]{joaoant@gmail.com}

\author[orcid=0000-0002-9269-8287]{Guilherme Limberg}
\affiliation{Kavli Institute for Cosmological Physics, University of Chicago, 5640 S. Ellis Avenue, Chicago, IL 60637, USA}
\affiliation{Department of Astronomy \& Astrophysics, University of Chicago, 5640 S. Ellis Avenue, Chicago, IL 60637, USA}
\email[]{limberg@uchicago.edu}

\author[orcid=0000-0002-9191-5972]{Lais Borbolato}
\affiliation{Instituto de Astronomia, Geof\'{\i}sica e Ci\^encias Atmosf\'ericas, Universidade de S\~ao Paulo, S\~ao Paulo, SP 05508-090, Brazil.}
\email[]{laisborbolato@usp.br}

\author[0000-0003-3922-7336]{Chervin F. P. Laporte}
\affiliation{LIRA, Observatoire de Paris, Université PSL, Sorbonne Université, Université Paris Cité, CY Cergy Paris Université, CNRS, 92190 Meudon, France}
\affiliation{Institut de Ciències del Cosmos (ICCUB), Universitat de Barcelona, Martí i Franquès 1, E-08028 Barcelona, Spain}
\affiliation{Kavli IPMU (WPI), UTIAS, The University of Tokyo, Kashiwa, Chiba 277-8583, Japan}
\email[]{chervin.laporte@gmail.com}

\author[0000-0001-5017-7021]{Zhao-Yu Li}
\affiliation{Department of Astronomy, School of Physics and Astronomy, Shanghai Jiao Tong University, 800 Dongchuan Road, Shanghai, 200240, China} 
\email[]{lizy.astro@sjtu.edu.cn}

\author[orcid=0000-0001-6880-4468]{Reinaldo Santos-Lima}
\affiliation{Instituto de Astronomia, Geof\'{\i}sica e Ci\^encias Atmosf\'ericas, Universidade de S\~ao Paulo, S\~ao Paulo, SP 05508-090, Brazil.} 
\email[]{reinaldo.lima@iag.usp.br}

\author[orcid=0000-0001-7902-0116]{Victor P. Debattista}
\affiliation{Jeremiah Horrocks Institute, University of Lancashire, Preston,
  PR1 2HE, UK.} 
\email[]{vpdebattista@gmail.com}

\author[orcid=0000-0002-9191-5972]{Marcelo C. Vicentin}
\affiliation{Instituto de Astronomia, Geof\'{\i}sica e Ci\^encias Atmosf\'ericas, Universidade de S\~ao Paulo, S\~ao Paulo, SP 05508-090, Brazil.} 
\email[]{marcelovicentin17@gmail.com}

\begin{abstract}
Recent evidence points for an intermediate-mass black hole (IMBH) in the most massive globular cluster of the Milky Way, $\omega$Centauri. This globular cluster is commonly associated with the Gaia-Sausage/Enceladus (GSE) dwarf -- 
the 
last major merger event of our Galaxy -- as its nuclear star cluster remnant. Motivated by this, we investigate whether the interaction between a bright GSE-like dwarf galaxy with stellar mass within $10^{8.5}$--$10^{9.7}$ solar masses and a Milky Way-like host can either suppress or enhance the growth of a central black hole (BH) during the merger process.
To this end, we perform high-resolution $N$-body $+$ smoothed particle hydrodynamics (SPH) simulations with the \code{ChaNGa} framework and a modified, angular-momentum-aware Bondi-Hoyle accretion prescription, considering both isolated dwarf galaxies and merger simulations under different stellar feedback efficiencies. The merger simulations show that the interaction does not halt BH accretion; instead, tidal interactions and gas inflows during pericentric passages strongly enhance accretion, leading to growth in the BH mass by more than two orders of magnitude in both low- and high-feedback models. The enhanced accretion is accompanied by a suppression of star formation in the dwarf galaxy. Our results suggest that mergers similar to the GSE event can efficiently fuel IMBH growth in the satellite galaxy rather than suppress it. This may be in tension with the estimated low mass of the proposed IMBH in $\omega$Centauri’s, but it could provide a mechanism for triggering dwarf active galactic nuclei activity.
\end{abstract}



\section{Introduction} 

Understanding the driving physical processes behind the formation of galaxies and their co-evolution with their central black holes (BHs) is a major goal of Astrophysics \citep[e.g.,][]{Lynden_1969, WhiteRees1978, Kauffmann_2000, Croton_2006, Kormendy2013blackHoles, Inayoshi2020smbh}. 
In this context, the local demographics of BHs residing in the inner regions of massive galaxies, including the Milky Way (MW) itself \citep{EckartGenzel1997sgrA*, Ghez1998sgrA*}, have revealed scaling relations between central BH mass (\mbh) and host properties such as luminosity or stellar mass (\mstar), which are regulated by active galactic nuclei (AGN) activity \citep{DiMatteo2005AGNfeedback}. However, the role of central BHs and associated AGN feedback in the evolution of low-mass ($\log{\mstar/\msun} \lesssim 9.5$) dwarf galaxies, i.e., with \mstar about that of the Large Magellanic Cloud (LMC) and lower, has remained elusive (see \citealt{silk2017_AGNinDwarfs} for an initial discussion).

It is generally thought that AGN helps shut down star formation in massive galaxies, while in low-mass systems, it is supernova feedback that suppresses it \citep[e.g., the early exploration by][]{Larson1974_SNfeedbackGals}. However, with the advent of wide and deep spectroscopic surveys such as the Sloan Digital Sky Survey \citep[SDSS;][]{sdssYork} and the significant number of dwarf AGN discoveries \citep[e.g.,][]{Reines2013bhs}, this view has been challenged. More recently, data from the Dark Energy Spectroscopic Instrument survey has provided the largest homogeneous collection of dwarf AGNs with more than 2,000 detections \citep{Pucha2025_dwarfAGNsDESI}.


Despite these large sample numbers, extracting useful statistics, such as BH scaling relations, for dwarf AGNs still has limitations. For example, the traditional optical emission-line ``Baldwin--Phillips--Terlevich'' BPT diagnostic diagram \citep{BPTdiagram1981} has its limitations in the dwarf-galaxy regime due to metallicity effects, as real AGNs can masquerade as star-forming systems \citep[][and \citealt{Mezcua2017review} for a review]{Groves2006metalpoorAGN}. This occurs because lower-mass galaxies are more metal-poor than their massive counterparts following the luminosity/\mstar--metallicity relation \citep[][and \citealt{kirby2013mzr} for dwarfs specifically]{Tremonti2004mzr, Mannucci2010mzr}. Therefore, the above-mentioned optical spectroscopic samples systematically underestimate the real active fraction of dwarf galaxies, as demonstrated in \citet{WasleskeBaldassare2024dwarfAGNs} using various AGN selection methods.

Beyond just observational aspects, modeling assumptions also hinder the reliability of dwarf AGN properties extracted from current data. \citet{burke2025dwarfAGN} compiled X-ray fluxes for nearby galaxies to estimate a central BH occupation fraction that reaches $\gtrsim$90\% at $\mstar = 10^8\,\msun$ and 100\% at $\mstar = 10^9\,\msun$. Despite similar underlying data and statistical framework, \citet{zou2025dwarfAGN} reached dramatically different conclusions, with occupation fractions of $\sim$20\% and $\sim$50\% for dwarf galaxies at $\mstar = 10^8\,\msun$ and $\mstar = 10^9\,\msun$, respectively. According to these authors, the discrepancy appears to be due to different assumptions regarding the accretion-rate distribution functions adopted. These unsolved issues illustrate current uncertainties in co-evolution between dwarf galaxies and their central BHs.

The ever-increasing number of known AGNs in bright dwarf galaxies ($\mstar \geq 10^8\,\msun$) combined with their non-negligible BH occupation fraction certainly demands that the community scrutinize the possible effects of these central BHs in the evolution of their host systems. Not just that, but the stellar population in the MW's halo is also composed of debris from disrupted dwarfs \citep[e.g.,][]{Helmi2020}, as predicted by hierarchical assembly within the cold dark matter framework \citep{WhiteRees1978, DEFW_1985, White1991cdm, Johnston1998, BullockJohnston2005}. Thanks to the \textit{Gaia} space mission \citep{GaiaMission, gaiadr1, gaiadr2, Gaia2021_EDR3, GaiaDR3_2023}, it has become clear that the Galactic halo is dominated by a single massive merger that happened about 10\,Gyr ago \citep[e.g.][]{gallart2019, Bonaca2020, Montalban_2021, laporte2026ocen, Orkney_2026} with a galaxy of similar mass to the LMC (see \citealt{Lane2023_GSEmass, Orkney_2026} for a critical discussion, and \citealt{DeasonBelokurov2024rev} for a review), dubbed `Sausage' or `Enceladus', hereby ``\textit{Gaia}-Sausage/Enceladus'' \citep[GSE;][]{Belokurov_2018, Haywood2018, Helmi_2018}. Following the occupation fractions reported by \citet{burke2025dwarfAGN} and \citet{zou2025dwarfAGN}, GSE itself could plausibly have hosted a central BH.


With the consolidation of the idea that GSE was the last significant merger experienced by the MW, some early works dynamically associated the most massive star cluster in the MW, NGC~5139 or $\omega$Centauri (\ocen), with it \citep[][but see also \citealt{Myeong_2019}]{Massari_2019, Callingham_2022}. In fact, \ocen has long been understood to be the stripped nuclear star cluster of a disrupted dwarf galaxy \citep{Lee_1999, hilker2000ocen, Bekki_2003}. More recently, \citet{laporte2026ocen} showed, for the first time, dynamical evidence supporting this association; they proposed that bar resonances can dislodge \ocen from the GSE debris and presented the first tentative evidence for a high-metallicity corridor connecting GSE to \ocen as expected from predictions from bar models in cosmological simulations \citep{tomlinson+2026}. \citet{mannu+2026} further strenghted this scenario by showing that the 3:2 retrograde bar resonance can dislodge \ocen from the GSE and by identifying N-rich stars compatible with \ocen at the low-energy end of the GSE and in the expected GSE--\ocen corridor \citep[see also][]{souza2026}; unlike the 1:1 resonance, which requires a very slow bar \citep{dillamore2026ocen}, the 3:2 resonance is natural for any bar model consistent with present-day constraints \citep{portail+2017}, making it very likely that \ocen was affected by a bar resonance during the bar's 8--10\,Gyr lifetime.

Another compelling argument for this connection has been formulated by \citet{Limberg_2022}, who demonstrated that, except for GSE, none of the known MW accretion events were massive enough to host a nuclear star cluster as massive as \ocen based on empirical scaling relations \citep[e.g.,][]{Neumayer2020reviewNSC}. \citet{laporte2026ocen} have recently arrived at a similar conclusion based on age--metallicity relations of \ocen's stellar populations and those of the GSE's globular clusters.

\citet{Haberle_2024} has provided evidence from fast-moving stars within \ocen that this star cluster hosts an intermediate-mass BH (IMBH) with $\mbh \gtrsim 8000\,\msun$ (see \citealt{noyola2008ocen} for an earlier proposition). This candidate IMBH in \ocen is one of not even a handful known in the $10^2 < \mbh/\msun < 10^5$ range \citep[][for a compilation]{Greene_2020}. \citet{Limberg_2024} then confirmed that the IMBH mass reported by \citet{Haberle_2024} does follow aforementioned BH scaling relations with \ocen and GSE properties. While previous theoretical works have explored different facets of dwarf-galaxy and central BH co-evolution, including in a cosmological context \citep{sharma2020romulus, koudmani2021sims, koudmani2022sims, koudmani2025sims, haidar2022sims, ArjonaGalvez2024}, none of them were fine-tuned for the GSE merger event with the MW. To understand how BH accretion history behaves in a GSE-mass galaxy as it merges and interacts with a MW-like host and to test if the properties of \ocen's IMBH are consistent with this scenario, we must simulate specific conditions such as host mass and orbital properties.

In this work, we expand the Gaia–EncelAdus–Sausage Timing, chemistRy and Orbit simulation suite (\texttt{GASTRO}; \citeauthor{Amarante_2022} \citeyear{Amarante_2022}, \citeauthor{Amarante_2025_GASTROII} \citeyear{Amarante_2025_GASTROII}) and explore the specific scenario where the BH-host GSE-like dwarf is also the satellite of a simulated MW analog. This work is organized as follows. Section~\ref{sec:numerical_methods} introduces the governing equations and physical background of our simulations. Section~\ref{sec:ic_setup} describes the initial conditions and simulation parameters. In Section~\ref{sec:results}, we detail and discuss the results of our simulations. In Section~\ref{sec:discussions}, we discuss our results and contextualize them with other works. Lastly, Section~\ref{sec:conclusion} summarizes the main findings and proposes next steps.

\section{Numerical methods} \label{sec:numerical_methods}

The simulation setup is identical to the \code{GASTRO} simulation library \citep{Amarante_2022, Amarante_2025_GASTROII} used to reproduce GSE-like accretion events
onto MW analogs \citep[see also][]{Borbolato_2026}, except for the inclusion of a central BH in the GSE dwarf. Nevertheless, we provide a thorough description of the input physics below to guarantee that the current paper is self-sufficient.

We run the experiments with the $N$-body + smoothed particle hydrodynamics (SPH) code \code{ChaNGa} \citep{Jetley_2008, Jetley_2010, Menon_2015}, a parallel simulation framework implemented in \code{Charm++} \citep{Kale_1993}. \code{ChaNGa} is based off the well-established tree+SPH code \code{GASOLINE} \citep{Wadsley_2004}, the latter being used to run \texttt{GASTRO}.

\subsection{\code{ChaNGa} Input Physics}\label{sec:inputphysics}

A pressure floor on gas particles is set so that
\begin{equation}
    p_{\rm floor} = 3 G\epsilon^2\rho^2,
\end{equation}
where $G$ is the universal gravitational constant, $\epsilon$ is the softening length, and $\rho$ is the gas particle's density. Star formation begins where the gas temperature drops below 15,000\,K, and the density exceeds $1\,\rm{cm}^{-3}$ \citep{Stinson_2006}. We consider the supernova feedback to be modeled by a blast-wave implementation \citep{Stinson_2006}, where we choose between $20\%$ or $80\%$ of $10^{51}\ \rm{erg}$ per supernova as thermal energy injected into the interstellar medium (models named with \code{20} and \code{80}, respectively). Exploring different feedback efficiencies is particularly important in our study because supernova feedback can regulate both the gas supply available for BH accretion and the level of star formation in the central regions of the dwarf galaxy. We consider the metallicity-dependent cooling for gas particles as described in \citet{Shen_2010}. We also include turbulent diffusion of the gas \citep{Shen_2010}, which allows for local mixing of metals and thermal energy between gas particles. 

We choose the base time step of $\Delta t = 10\ \rm{Myr}$ and refine it to account for faster dynamical changes following 
\begin{equation}
    \delta t = \Delta t/2^2 < \nu\sqrt{\epsilon/a_g},
\end{equation}
where $a_g$ is the acceleration at a gas particle's position and the refinement parameter is $\nu = 0.175$. The time step is also constrained by the Courant condition, where 
\begin{equation} 
\delta t = \frac{\nu_{\rm Courant} h}{(1 + \alpha_{\rm visc})c + \beta_{\rm visc} \mu_{\rm max}},
\end{equation}
where the Courant number for the SPH criterion is $\nu_{\rm Courant} = 0.4$, $h$ is the smoothing length calculated over the nearest 32 particles, $\alpha_{\rm visc} = 1$ and $\beta_{\rm visc} = 2$ are the linear and quadratic viscosity coefficients, and $\mu_{\rm max}$ is the maximum artificial-viscosity signal velocity between neighboring gas particles \citep{Wadsley_2004}. This ensures that gas particles do not move more than one smoothing length per time step. Gravitational forces are computed using a Barnes–Hut tree algorithm with an opening angle $\theta = 0.7$, which sets the criterion for approximating distant particle groups by their multipole expansion.

\subsection{Black Hole Accretion Physics}

In our simulations, the black hole accretes mass from, and injects feedback energy into, its 64 nearest neighboring gas particles. We adopt this expanded 64-particle interaction sphere, rather than a standard 32-particle gas kernel, to prevent artificial gas blowouts and ensure numerical stability within the shallow potential wells of dwarf galaxies. The energy released from the accretion is distributed spherically, weighted by the Wendland C4 smoothing kernel (meaning gas closer to the BH receives a higher fraction of the energy). To mimic the continuous transfer of energy from the central BH to the surrounding gas, particles that receive feedback energy are prevented from cooling for a period equal to the BH timestep. The amount of energy coupled to the surrounding gas particles is given by
\begin{equation}
    E = \eta_{\rm rad} \epsilon_{\rm{f}} \dot{M}c^2 dt. 
    \label{eq:accretion_energy}
\end{equation}
Here, $\eta_{\rm rad}$ is the radiative efficiency, which is assumed to be the standard value of $10\%$ \citep{Shakura_1973, Zhang_2017}, $\epsilon_{\rm{f}} = 0.005$ is the efficiency at which radiated energy is transferred to the surrounding gas, $\dot{M}$ is the BH accretion rate, $c$ is the speed of light, and $dt$ is the timestep value.

The accretion rate is estimated via a modified Bondi-Hoyle prescription \citep{Bondi_1952, Tremmel_2017}. Using the same energy balance as used in the derivation of the Bondi-Hoyle accretion model, the code recalculates the accretion radius, including the effect of angular momentum support based on the resolved dynamics of the gas in the simulation (as detailed in \citeauthor{Tremmel_2017}, \citeyear{Tremmel_2017}). We also consider a density-dependent boost factor to account for the unresolved nature of the gas near the central BH \citep{Booth_Schaye_2009}, resulting in

\begin{equation}
\dot{M} = \left( \frac{n}{n_{\rm th,*}} \right)^\beta
\begin{dcases}
    \frac{\pi (GM_{\rm BH})^2 \rho}{(v_{\rm bulk}^2 + c_s^2)^{3/2}} & \text{if } {v}_{\rm bulk} > v_\phi, \\
    \\
    \frac{\pi (GM_{\rm BH})^2 \rho c_s}{(v_\phi^2 + c_s^2)^2} & \text{if } {v}_{\rm bulk} < {v}_\theta.
\end{dcases}
\label{eq:macc_rate}
\end{equation}
Here \(v_\phi\) is the tangential velocity of the gas relative to the BH, representing the rotational component of the gas motion, while \(v_{\rm bulk}\) is the bulk velocity of the surrounding gas relative to the BH, representing the overall relative motion of the gas. The comparison between these quantities provides an estimate of the dynamical state of the gas around the BH; gas with \(v_\phi \gg v_{\rm bulk}\) is primarily rotationally supported, whereas gas with \(v_{\rm bulk} \gtrsim v_\phi\) has a stronger radial or disordered component, indicating more efficient inflow toward the BH. $\rho$ is the density of the gas, $c_s$ is the sound speed of the gas, $M_{\rm BH}$ is the mass of the BH, $n_{\rm th, *}$ is the number density threshold for star formation, $n$ is the local gas number density and $\beta$ is the boost factor. For the simulations considered in this paper, we considered $\epsilon_{\rm f} = 0.005$ and $\beta = 2$. A more detailed analysis of the parameters $\epsilon_{\rm f}$ and $\beta$ was presented in \cite{Tremmel_2017}. Although \cite{Tremmel_2017} identified $\beta = 2$ and $\epsilon_{\rm f} = 0.02$ as the best-fitting parameters for their cosmological simulations, we do not adopt those values directly because our simulations focus on an isolated dwarf galaxy merger setup. Instead, we performed a dedicated parameter tuning procedure, adjusting these values until sustained black hole growth was obtained in our simulations. The values adopted in this work therefore differ from the best-fitting parameters reported by \cite{Tremmel_2017}, while remaining within the physically plausible range explored by the authors. It should be noted, however, that these parameters are part of a subgrid prescription designed to model unresolved physical processes, and their adopted values may depend on the specific numerical setup and calibration strategy.


\subsection{Black Hole Dynamics}
Dynamical friction is responsible for the decay of the orbits of massive BHs towards the centers of massive galaxies \citep{Governato_1994, Kazantzidis_2005}. This effect is usually unresolved in cosmological simulations due to numerical noise and low gravitational force resolution \citep{Tremmel_2015, Tremmel_2017, Genina_2024}. As in \citet{Tremmel_2015}, we assume that the simulation resolves the gravitational wake produced by the BH on scales larger than the softening length, but it can not resolve the wake inside it. Therefore, an extra acceleration that accounts only for the unresolved portion of the dynamical friction is added following Chandrasekhar's formalism \citep{Chandrasekhar1943dynfriction, Tremmel_2015}. At each time step, the local density and velocity distribution are estimated from the nearest particles surrounding the BH, and an additional acceleration opposing the BH's motion is computed from the unresolved dynamical friction below the gravitational softening scale. This correction is applied only when the relevant dynamical-friction scales are not already resolved by the simulation.

\section{Initial conditions setup}
\label{sec:ic_setup}

\subsection{Host Galaxy Setup}

The host galaxy's initial condition (IC) is set such that it will end up as a MW-like galaxy after several Gyr of evolution. We adopt the ICs that have been extensively used in previous studies that reproduce the chemo-geometrical properties of the MW's disk \citep{Clarke_2019, Beraldo_Silva_2020, Amarante_2025_GASTROII}, the kinematics of the \textit{in situ} halo \citep{Amarante_2020, Borbolato_2026}, the co-formation of the $\alpha$-rich and $\alpha$-poor disk populations \citep{Beraldo_Silva_2021, Amarante_2025_GASTROII}, and the chemistry of the bulge \citep{Debattista_2023}.

The IC is set up with a spherical hot corona embedded in a dark matter (DM) halo with a Navarro-Frenk-White \citep{Navarro_1997} cuspy density profile. For the DM, we consider a virial radius $r_{\rm 200} = 200\ \rm{kpc}$, central density $2.4\times 10^6 M_\odot \, \rm{kpc}^{-3}$, mass of $10^{12}\,M_\odot$, and scale radius of $30\,\rm{kpc}$. For our experiments, we increased the gas and DM mass resolution by a factor of 2 and 4, respectively, compared to the original \texttt{GASTRO} models. DM particles have two different masses: $2.5\times10^5 M_{\odot}$ if $r < 200 \ \rm{kpc}$ and $8.85 \times 10^5 M_\odot$ if $r > 200 \ \rm{kpc}$. The gas particles follow the radial density distribution of the DM particles and have an initial net rotation with spin of $\lambda = 0.065$ \citep{Bullock_2001} to allow it to form a disk. The initial mass of the gas corona is $1.4 \times 10^{11} M_\odot$. Each gas particle's mass is set to $M_{\rm gas} = 6.96 \times 10^4 M_\odot$. The softening length used for the DM and gas is 100\,pc and 50\,pc, respectively.  Star particles form self-consistently from cold gas, inheriting both the softening length and chemical properties of their parent gas particles.

\subsection{Dwarf Galaxy Setup}

The dwarf galaxy model adopted in this work is the same as that from \citet{Amarante_2025_GASTROII}. It is generated using the \code{GALACTICS} code \citep{Kuijken_1995, Widrow_2005, Widrow_2008,deg+2019}. For the simulations that consider the merger of the dwarf and the MW-like host galaxy, we consider a DM halo and gas disk with  $4\times10^5$ and $2 \times 10^4$ particles, respectively. The DM particles are distributed following an NFW density profile with a virial radius of 91\,kpc, a truncation radius of 50\,kpc, and a truncation width of 5\,kpc (see \citealt{Widrow_2008} for details). 
The gas density profile follows an exponential profile with a scale length of 1\,kpc, and the initial gas mass of the dwarf is set to $1.4 \times 10^9 M_{\odot}$. It starts with a kinematic temperature of 40,000 K, which keeps the gas disk in hydrostatic equilibrium. 

For the isolated dwarf galaxy simulations, we adopt a substantially higher mass resolution, increasing the number of DM and gas particles by factors of 30 and 7.5, respectively, relative to the merger simulations. The reduced computational cost of evolving the dwarf in isolation enables this increase in resolution. A higher mass resolution also permits the use of lower-mass black hole seeds, as the black hole particle mass is no longer constrained by the comparatively large gas particle masses required in the merger runs. This allows us to assess the impact of numerical resolution on the inferred accretion history by comparing the isolated and merger simulations.

\subsection{Black Hole Setup}

We evolve the dwarf galaxy in isolation for $0.2\ \rm{Gyr}$, equivalent to redshift\footnote{Assuming a concordance $\Lambda$CDM cosmology with $H_0 = 70 \rm \ km/s/kpc$, $\Omega_m = 0.3$ and $\Omega_\Lambda = 0.7$.}  $z \sim 18$, before introducing the MW particles and inserting the BH. This allows the dwarf to relax from its initial conditions, decreasing any non-physical effects on the subsequent BH accretion. Then, to satisfy the conditions where $M_{\rm BH}$ is greater than the individual mass of each gas particle, we introduce a BH of mass $10^5 M_\odot$ in the merger runs, and $8 \times 10^3\ M_\odot$ in the isolated runs, following Table~\ref{tab:simulations}, in the center-of-mass coordinates of the halo of the dwarf galaxy, with the orbital speed calculated using the mass-weighted average velocity of the 32 closest star and DM particles to the center. We note that the inferred BH mass in \ocen is of order ${\rm \sim}10^4 \,M_\odot$, making our adopted seed mass comparable to observational estimates. Although the exact seed mass is constrained by the requirement $M_{\rm BH} > M_{\rm gas}$ at our simulation resolution, our primary goal is not to reproduce the BH mass of \ocen precisely. Instead, we focus on the BH growth history and on determining whether the merger-driven evolution can sustain or suppress central BH accretion. We verified that varying the seed mass between $10^4$ and $10^5\,M_\odot$ in isolated simulations produces nearly identical accretion histories and growth rates, indicating that our results are not sensitive to the precise seed mass within this range.


An important caveat is that the initial BH mass adopted in our simulations is below the mass scale required to fully resolve the BH dynamical evolution. Following \citet{Tremmel_2015}, the BH mass should be larger than the background particle mass to ensure a reliable treatment of its dynamical evolution; in our case, this limitation is primarily set by the DM particle mass. However, before and at the time of the merger interaction, which is the main focus of this work, the BH has grown above this resolution threshold, making its dynamical evolution better resolved during the relevant phase. To further verify that the BH remains dynamically associated with the dwarf galaxy, we analyze its distance from the galaxy center throughout the interaction and confirm that it remains closely bound to the dwarf galaxy center until the stripping process begins.

\subsection{Model Impact Parameters}\label{sec:impactparam}

The dwarf is set at $(X, Y, Z) = (-6.17, -191.21, 50.18)$ kpc relative to the center of mass of the simulated MW-like host with velocities $(V_x, V_y, V_z) =
(-29.71, 20.36, -5.38)$\,\kms. Its orbital plane inclination relative to the host galaxy’s midplane is set to $15^{\circ}$. These parameters mimic model \code{D1}, described in \citet{Amarante_2022}, and were motivated by MW$+$GSE \textit{N}-body models from \citet{naidu2021}. The corresponding retrograde orbit has an initial circularity of $\eta \equiv |L_z| / L_c(E_n) = 0.3$, where $L_z$ is the $z$-component of the angular momentum and $L_c(E_n)$ is the angular momentum of a circular orbit with the same energy $E_n$ in the Galactic plane. 

\section{Results}
\label{sec:results}
In this section, we present results from six simulations, two of them modeling the GSE-like dwarf galaxy in isolation with a central BH (\code{ISOBH20} and \code{ISOBH80}), two modeling it in isolation without a BH (\code{ISO20} and \code{ISO80}), and two incorporating a merger event with a central BH in the simulated dwarf galaxy (\code{M20BH} and \code{M80BH}). The numerical suffixes (\code{20} and \code{80}) indicate the assumed fraction of stellar feedback energy—specifically, $20\%$ and $80\%$ of $10^{51}$\,erg per supernova, respectively.  All the models evolve for $5$\,Gyr. This is enough to study the effect of an IMBH in an isolated dwarf, and also in the merger scenario, as the final pericentric passage occurs at $t \sim3.2$\,Gyr after the simulation starts. We also compare our results to simulations without a central BH, named \code{M20} and \code{M80}, presented in \cite{Amarante_2022}.

\begin{deluxetable*}{cccccc}
\tablecaption{Summary of the simulations analyzed in this work. The numerical suffixes 20 and 80 correspond to feedback energies of 20\% and 80\% of $10^{51}$\,erg per supernova (SN), respectively. All simulations adopt a gravitational softening length of $0.05$\,kpc and are evolved for 5\,Gyr. \label{tab:simulations}}
\tablehead{
\colhead{Simulation} & \colhead{Type} & \colhead{SN Feedback Energy} & \colhead{DM Mass Resolution } & \colhead{Gas Mass Resolution}  & \colhead{BH Initial Mass} \\
\colhead{} & \colhead{} & \colhead{($10^{51}\,\mathrm{erg}$)} & \colhead{(\msun)} & \colhead{(\msun)} & \colhead{($M_\odot$)}
}
\startdata
\texttt{ISO20}   & Isolated      & 20\% & $1.47\times10^4$ & $4.37\times10^3$ & N/A \\
\texttt{ISO80}   & Isolated      & 80\% & $1.47\times10^4$ & $4.37\times10^3$ & N/A \\
\texttt{ISOBH20} & Isolated + BH & 20\% & $1.47\times10^4$ & $4.37\times10^3$ & $8 \times 10^{3}$ \\
\texttt{ISOBH80} & Isolated + BH & 80\% & $1.47\times10^4$ & $4.37\times10^3$ & $8 \times 10^{3}$ \\
\texttt{M20BH}   & Merger + BH   & 20\% & $2.5\times10^5$ / $8.85\times10^5$ & $6.96 \times 10^4$ & $10^{5}$ \\
\texttt{M80BH}   & Merger + BH   & 80\% & $2.5\times10^5$ / $8.85\times10^5$ & $6.96 \times 10^4$ & $10^{5}$ \\
\enddata

\end{deluxetable*}

\subsection{Isolated dwarf galaxy comparison}
\label{sec:isodwarf}

In this section, we compare global parameters of the isolated dwarf galaxy with and without the presence of the central BH. Here, we do not account for the merger with the MW-like host; therefore, we analyze only simulation runs \code{ISO20}, \code{ISOBH20}, \code{ISO80}, and \code{ISOBH80}.

\begin{figure}
    \centering
    \includegraphics[width=\linewidth]{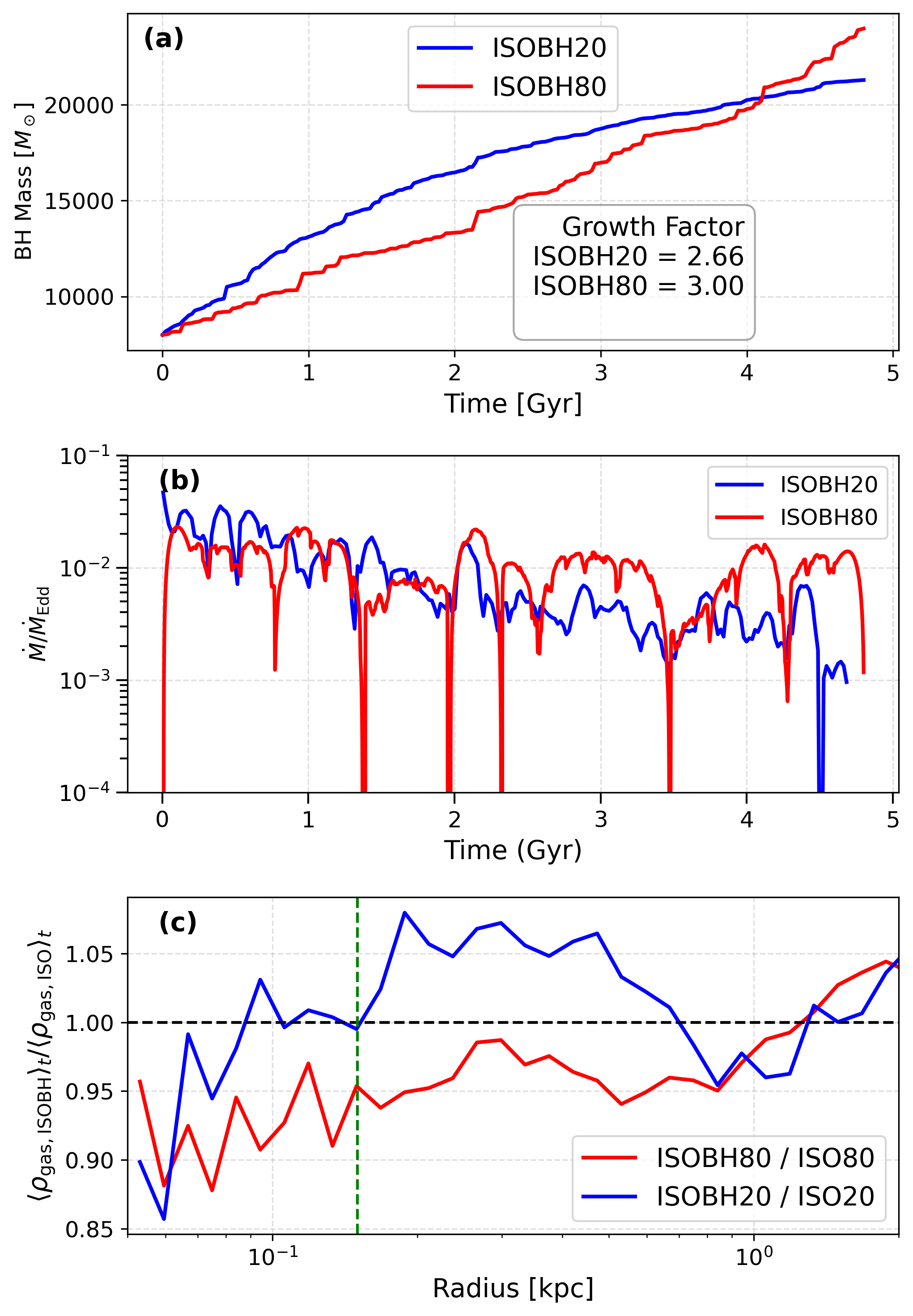}
    \caption{Panel (a): BH mass over the simulation time for the isolated simulations \code{ISOBH20} (blue) and \code{ISOBH80} (red). The growth factor ($M_{\rm BH}(t_{\rm{final}})/M_{\rm BH}(t_{\rm initial})$ of each BH is written explicitly on the plot. Panel (b): Accretion rate in units of the Eddington accretion rate for both simulations as a function of simulated time. The accretion rate is smoothed using a Savitzky–Golay filter with a window of 600 points, corresponding to approximately 0.125 Gyr. Panel (c): Radial profile of the mass density of the gas in the simulations with the presence of a BH (\code{ISOBH20} and \code{ISOBH80}), normalized by the ones without the BH (\code{ISO20} and \code{ISO80}), averaged across all the simulation time. The dashed vertical green line represents three times the softening length.
    }
    \label{fig:IsoBHGrowthAccRateRhoRadialProf}
\end{figure}

We start by investigating the BH growth alongside the mass accretion rate and the gas mass density radial profile averaged over the whole simulation time span in Figure~\ref{fig:IsoBHGrowthAccRateRhoRadialProf}. In the top panel (a), we observe a similar growth factor relative to the initial seed mass in both simulations, with values of ${\sim} 2.66$ for \code{ISOBH20} and ${\sim} 3.00$ for \code{ISOBH80}. A similar difficulty in growing central BHs in isolated dwarf-galaxy simulations had been previously reported by \citet{Hazenfratz2025dwarfBH} and \citet{Petersson_2025}. Although both simulations exhibit similar overall growth factors, \code{ISOBH20} grows more rapidly at early times, with its growth rate gradually declining thereafter. In contrast, \code{ISOBH80} maintains a nearly constant growth rate throughout the evolution.

In the middle panel (b) of Figure~\ref{fig:IsoBHGrowthAccRateRhoRadialProf}, we show the accretion rate normalized by the Eddington accretion rate,
\begin{equation}
    \dot{M}_{\rm Edd} = \frac{4 \pi G M_{\rm BH} m_p}{\eta_{\rm rad} c\sigma_T},
\end{equation}
as a function of simulated time. Here, $M_{\rm BH}$ is the BH mass, $m_p$ is the proton mass, and $\sigma_T = 6.652\times 10^{-25}\,\mathrm{cm^2}$ is the Thomson scattering cross-section. The trends observed here are consistent with the BH growth presented in panel (a), where \code{ISOBH80} exhibits a relatively steady accretion rate throughout its evolution and, in contrast, \code{ISOBH20} starts with a higher accretion rate, which gradually decreases over time
. This highlights that the assumed supernovae feedback does impact the BH accretion in our isolated runs.  

Lastly, in panel (c) of Figure~\ref{fig:IsoBHGrowthAccRateRhoRadialProf}, we show the gas density profile of the isolated simulations averaged over the whole simulation time; we notice that the presence of the BH decreases the inner mass density for radii $r \lesssim 0.15\ \rm{kpc}$ in both adopted supernova feedback values. \code{ISOBH80} presents an overall shallower gas density profile when compared to \code{ISO80} until $r \sim 1 \ \rm{kpc}$, while \code{ISOBH20} presents a region of higher density when compared to \code{ISO20}. However, these differences remain at the $\lesssim 5\%$ level, indicating that the BH simulations show only small differences in the overall gas distribution, consistent with the limited mass growth of the central BH.

\begin{figure}
    \centering
    \includegraphics[width=\linewidth]{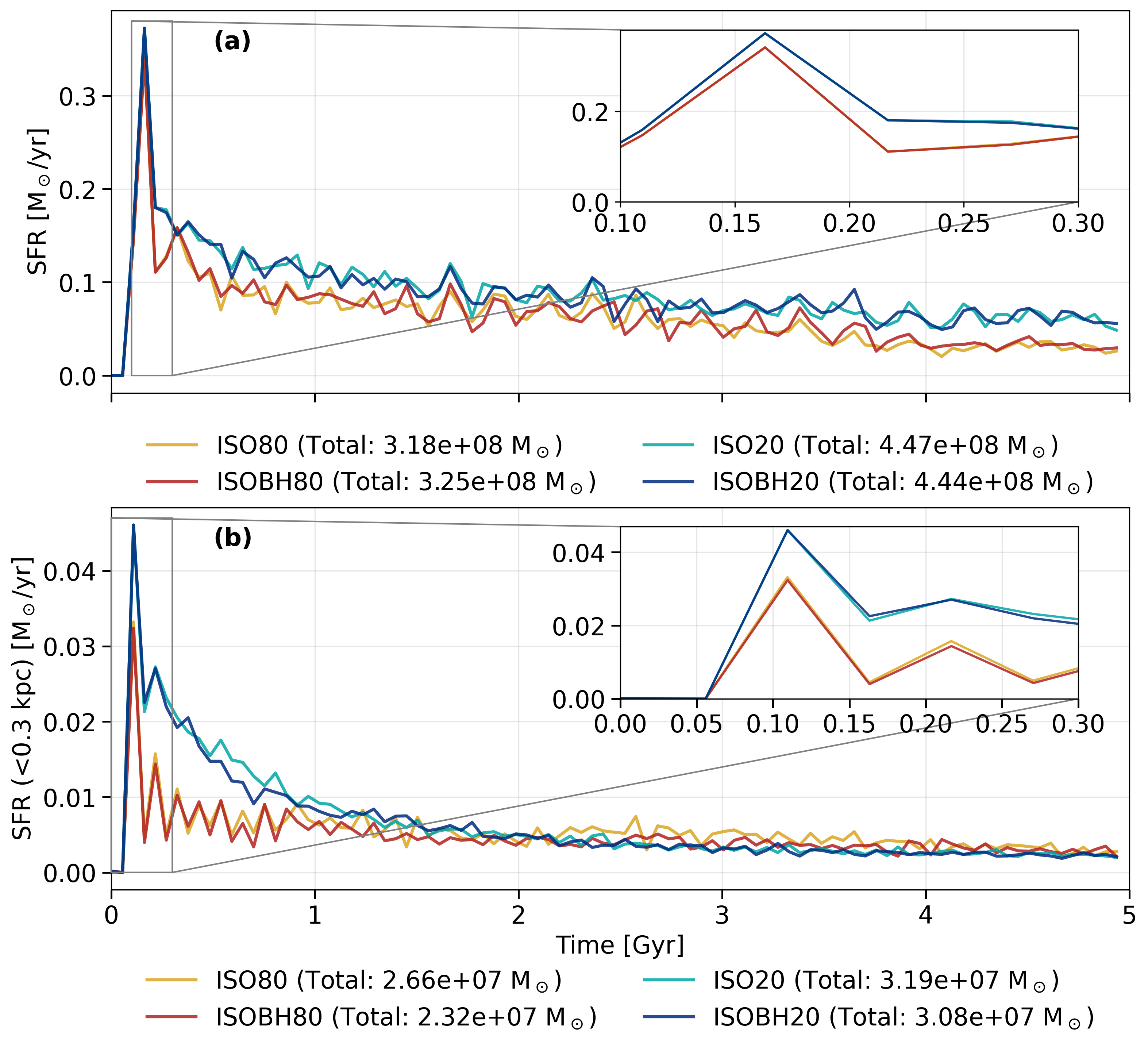}
    \caption{SFR as a function of simulation time for the isolated dwarf galaxy runs. Panel (a): Comparison of the global SFR for the four isolated simulations. The cool-colored curves correspond to the 20\% feedback efficiency models, while the warm-colored curves correspond to the 80\% feedback efficiency models. Within each color family, the two curves distinguish simulations with and without a central BH. The inset highlights the early peak in star formation, showing that the presence of the BH does not significantly affect the overall SFR. Panel (b): Same as panel (a), but restricted to stars formed within $r<0.3$ kpc, emphasizing the central region most affected by the BH.}
    \label{fig:SFR_ISO}
5\end{figure}

To investigate the impact of BH feedback on the temporal evolution of star formation, we examine the star formation rate (SFR) as a function of time for the isolated simulations in Figure~\ref{fig:SFR_ISO}. As shown in panel (a), the comparison between runs with and without a BH reveals that the presence of the compact object does not significantly alter the star formation history. It is possible to see a reduction in the SFR by a factor of $\sim$2$\times$ in the inner regions of the dwarf ($r < 0.3\, \mathrm{ kpc}$), which is commensurate with the reduction of the central gas density seen in Figure \ref{fig:IsoBHGrowthAccRateRhoRadialProf}. This suppression of SFR is driven by the feedback prescriptions instead of the central BH presence; models with weaker feedback sustain higher SFR for longer. The BH has minimal effect on the isolated dwarfs' SFR, only when considering the final total stellar mass, but it could be a stochasticity effect. The main driver in these isolated models is supernova feedback: lower feedback has higher SFR, which will translate to higher stellar mass and higher [Fe/H], as we will show later.

\begin{figure*}
    \centering
    \includegraphics[width=\linewidth]{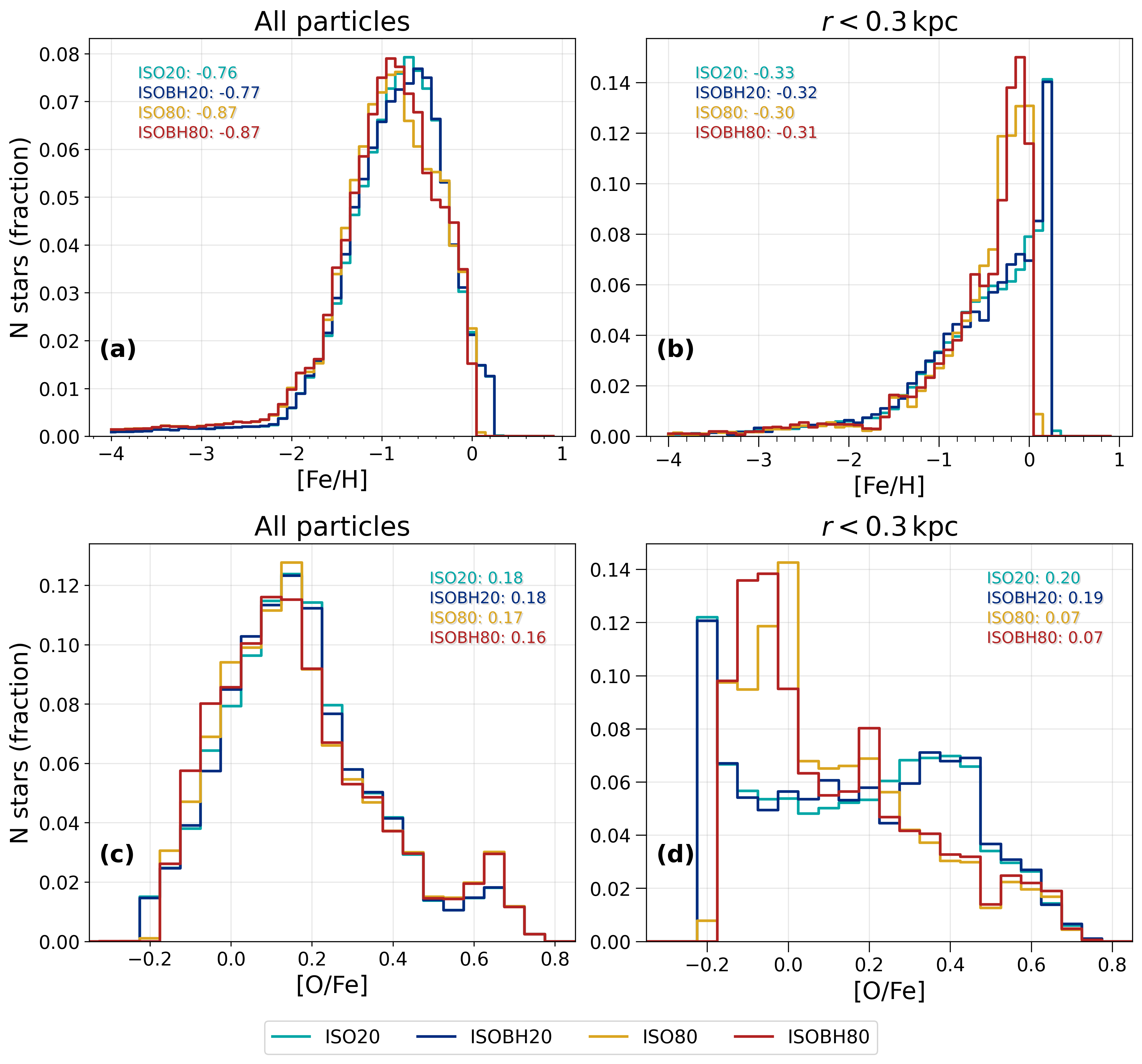}
     \caption{Stellar chemical abundance distributions for the dwarf galaxy simulations, comparing runs with and without a central BH. The top row displays [Fe/H] histograms for (a) all star particles and (b) stars restricted to the central $r < 0.3$ kpc. The bottom row shows the corresponding [O/Fe] distributions for (c) all star particles and (d) the central $r < 0.3$ kpc region. Cool-colored curves correspond to the 20\% feedback efficiency models, whereas warm-colored curves represent the 80\% feedback efficiency models. All histograms are normalized such that the total fraction sums to unity. The median abundance of each distribution is indicated within the corresponding panel.}
    \label{fig:ISOFeHDistributions}
\end{figure*}

Finally, to investigate the impact of the BH on the chemical evolution of the dwarf galaxy, we plot a histogram showcasing the distribution of [Fe/H] among the star particles in Figure~\ref{fig:ISOFeHDistributions} for the last snapshot of each simulation. Panel (a) shows the full stellar population, while panel (b) focuses on stars within the central 0.3\,kpc
. As can be seen from the distributions, the presence of the BH does not alter the overall metallicity distribution. The differences between simulations with and without a BH are subtle and largely confined to the very central regions. The same can be visualized for [O/Fe] in panels (c) and (d) of the same figure. These variations are small compared to the ${\sim}0.1$\,dex error floor even with high-resolution spectroscopy \citep{McWilliam_1997} and the typical spread in stellar abundances, indicating that, in practice, the impact of a central BH on the [Fe/H] and [O/Fe] distributions would be extremely difficult to detect. Despite not being the focus of this work, we verify that different supernova feedback prescriptions do have major impacts on the chemical enrichment of the dwarf galaxy's inner region ($r < 0.3\,{\rm kpc}$), as expected given its role in regulating star formation.

\subsection{Dynamical evolution and growth of the dwarf galaxy's central BH during the merger}

\begin{figure*}
    \centering
    \includegraphics[width=\linewidth]{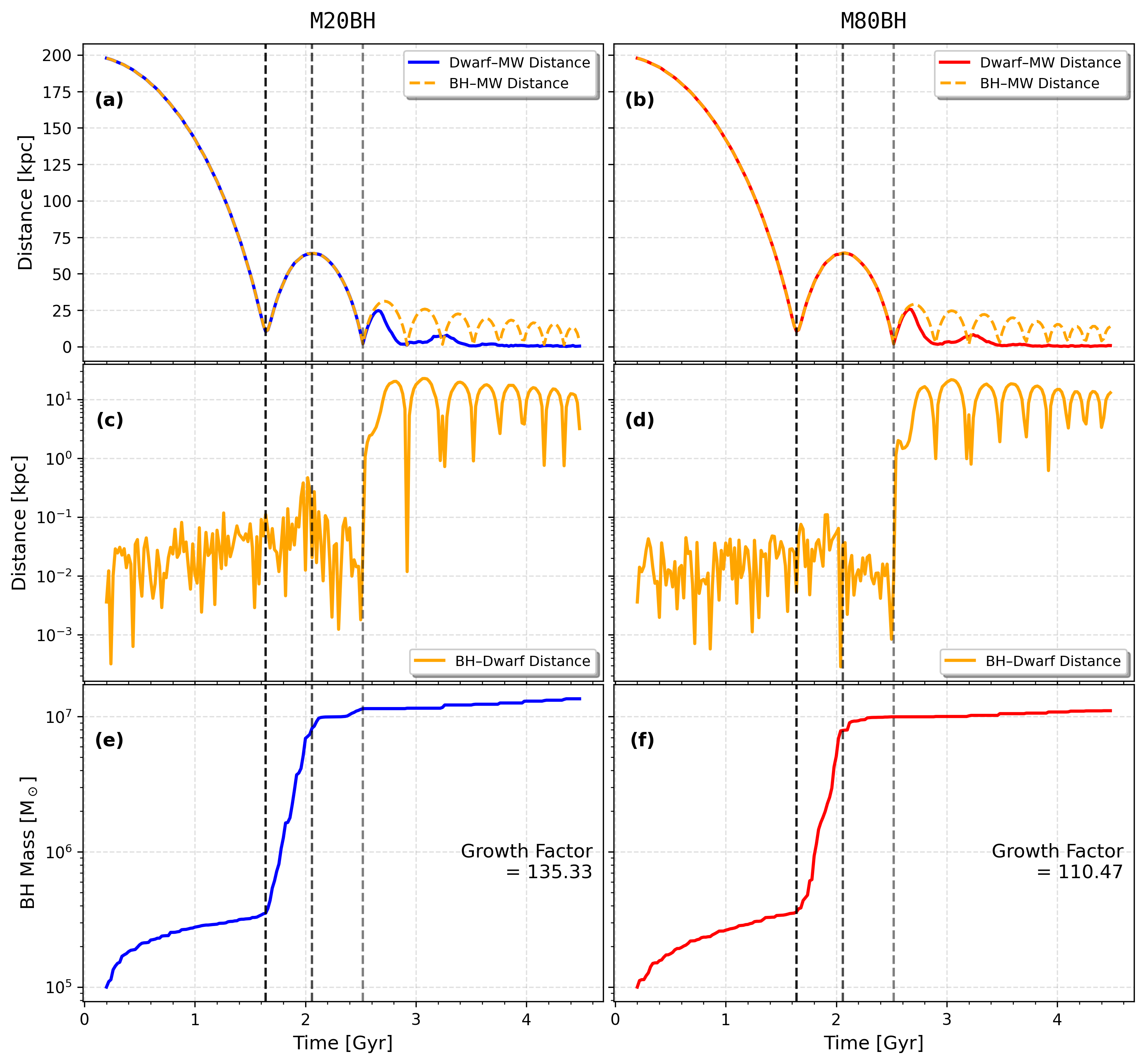}
    \caption{Evolution of the distance between the center of mass of the dwarf and the host galaxy for \code{M20BH} and \code{M80BH}, shown in panels (a) and (b), respectively. The dashed orange lines indicate the distance between the BH and the MW-like galaxy. Panels (c) and (d) show the distance between the dwarf galaxy's center of mass and the central BH. Panels (e) and (f) show the evolution of the BH mass for \code{M20BH} and \code{M80BH} over the course of the simulation. The vertical dashed lines mark the times of the first pericenter, the first apocenter, and the second pericenter.}
    \label{fig:BHposMassGrowthMergers}
\end{figure*}

In this Section, we look into the evolution of the central BH in the dwarf galaxy during the merger process with its host galaxy. As described in Section~\ref{sec:impactparam}, this attempts to mimic the GSE event in the MW. We first examine the evolution of the distance between the center of mass of the dwarf galaxy and that of the host galaxy, alongside the distance between the BH and the host galaxy’s center of mass. These evolutions are shown for both \code{M20BH} and \code{M80BH} in Figure~\ref{fig:BHposMassGrowthMergers} panels (a) and (b), respectively. Particular timestamps showcasing the first pericenter, first apocenter, and second pericenter epochs are depicted by vertical dashed lines. 

\begin{figure*}
    \centering
    \includegraphics[width=\linewidth]{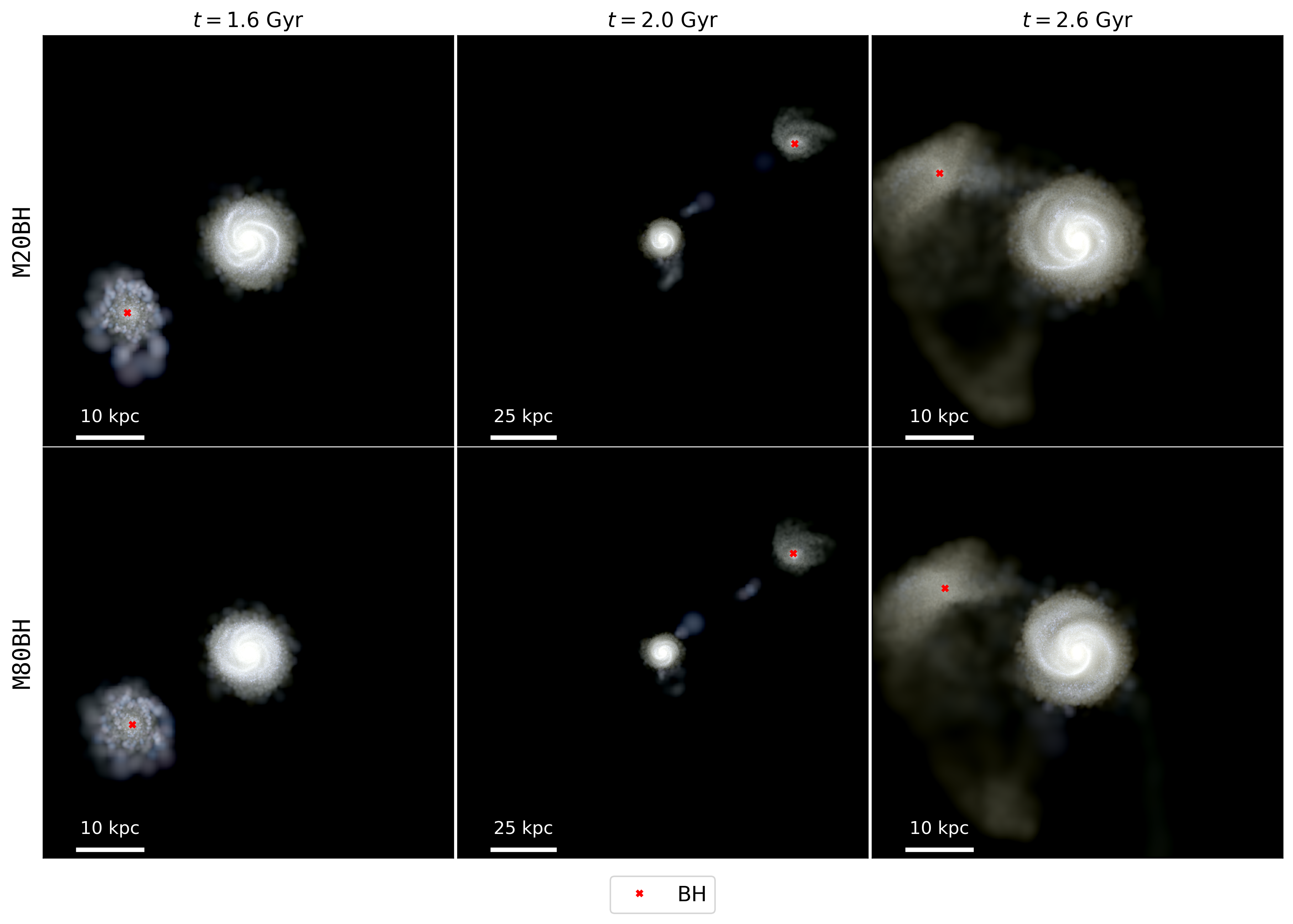}
    \caption{Rendering of the star particles and the merger for \code{M20BH} (top row) and \code{M80BH} (bottom row) for $t = 1.6$,  $2.0$, and $2.6$ Gyr, respectively. The red cross highlights the position of the BH.}
    \label{fig:stars_colormap}
\end{figure*}

The BH remains at the dwarf galaxy’s center of mass throughout most of the early orbital periods, as evidenced by panels (c) and (d). It only deviates significantly in the third pericentric passage, when much of the galaxy’s structure has already been altered by gravitational tidal effects, so the center of mass calculated does not represent the bulk motion of the inner star particles. Note that, after the third pericenter, since the dwarf has been fully disrupted, we no longer track its center of mass. This corroborates the expectation that, despite the initial BH mass being below the resolution threshold during the initial time of the simulation, the BH remains physically associated with the dwarf galaxy during the merger phase of interest, when its mass is sufficiently large for its dynamics to be reliably resolved. For illustration purposes, we show in Figure~\ref{fig:stars_colormap} a visualization of three snapshots of interest highlighting the position of the BH during the merger process.

To observe the effects of the merger on the BH growth, we plot the BH mass as a function of time for both \code{M20BH} and \code{M80BH} in Figure~\ref{fig:BHposMassGrowthMergers} panels (e) and (f), respectively. Prior to the first pericentric passage, the BH mass of \code{M20BH} and \code{M80BH} increases similarly to the isolated models. After this, the BH mass grows rapidly by ${\sim}2$ orders of magnitude in both models; this phase contributed the most to the overall growth over the course of the entire simulation, with final growth factors of $135$ for \code{M20BH} and $110$ for \code{M80BH}. These large growth factors are accompanied by accretion rates of $>$20\% of the Eddington ratio and we discuss their consequences below.

\begin{figure*}
    \centering
    \includegraphics[width=0.8\linewidth]{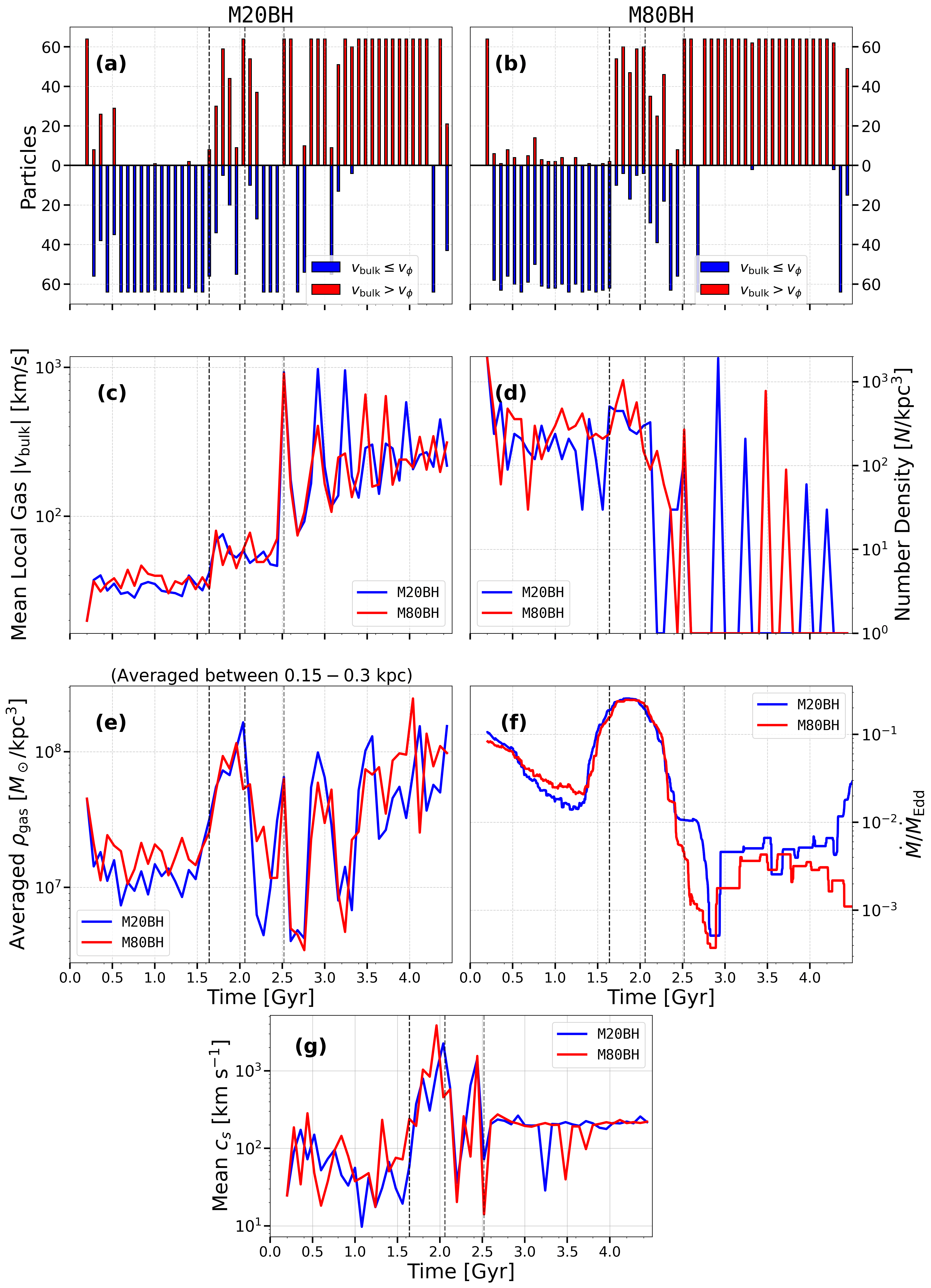}
        \caption{Time evolution of accretion-related parameters for \code{M20BH} and \code{M80BH}. Panels (a)-(b) show the distribution of gas particle bulk velocities relative to the local rotational velocity $v_\phi$ around the BH, separating particles with $v_{\rm bulk} \leq v_\phi$ and $v_{\rm bulk} > v_\phi$. Panel (c) presents the mean bulk velocity of the nearest gas neighbors of the BH. Panel (d) shows the evolution of the local gas number density within the vicinity of the BH. Panel (e) shows the time evolution of the spherically averaged gas density profile between $0.05$ and $0.5$ kpc. Panel (f) presents the BH accretion rate $\dot{M}_{\rm BH}$ in Eddington units. All quantities are extracted from snapshots spanning the simulation time and computed using the $N_{\rm smooth}=64$ nearest gas neighbors. Snapshots are spaced by $0.08\,{\rm Gyr}$ in time. The accretion rate shown in Panel (f) is smoothed using a moving average over $0.6\,{\rm Gyr}$. The vertical dashed lines in all panels denote the first pericenter, first apocenter, and second pericenter passages. Finally, Panel (g) showcases the mean local sound speed near the BH, considering the internal energy of the 64 particles.}
    \label{fig:AccretionParametersEvolution}
\end{figure*}

\subsection{Gas physics during major BH growth episodes} 

To better understand the origin of the BH accretion variability, we plot the time evolution of the gas properties in the vicinity of the BH, defined by the 64 nearest gas particles used in the accretion calculation, as described in Eq.(\ref{eq:macc_rate}), in Figure~\ref{fig:AccretionParametersEvolution} panels (a)-to-(f). The exception is the number density shown in panel (d), which is computed within a spherical region of radius 4 times the softening length centered on the BH. An initial sharp decline in both the particle number density and the averaged gas density (panels d and e) is observed following the introduction of the BH particle into the simulation at around $0.2$\,Gyr. At this stage, the gas represented by the 64 nearest neighbors surrounding the BH becomes subject to accretion and feedback processes that were absent before its insertion, leading to a depletion of the local gas reservoir either through direct accretion onto the BH or energy redistribution driven by feedback outflows. The large increase in BH mass up to $\sim 0.6$ Gyr with an accretion rate of $0.05 - 0.1 \dot{M}/\dot{M}_{\rm Edd}$ (panel f) indicates that, indeed, a fraction of the surrounding gas is accreted during this initial transient phase, after which the system relaxes toward a more steady configuration as seen in Figure~\ref{fig:BHposMassGrowthMergers} between ${\sim}0.6$\,Gyr and ${\sim} 1.6$\,Gyr.

Around the time we introduce the BH up to ${\sim} 0.2$\,Gyr, the kinematics of the gas in the vicinity of the BH transitions from a state in which $v_\phi < v_{\rm bulk}$ to one dominated by rotational support, where $v_\phi > v_{\rm bulk}$. This indicates that gas particles progressively settle into rotational motion around the BH, as shown in panels (a) and (b) of Figure~\ref{fig:AccretionParametersEvolution} for both the \code{M20BH} and \code{M80BH}. This configuration persists until $t \sim 1.6$ Gyr, when the dwarf galaxy reaches its first pericentric passage, as illustrated in Figure~\ref{fig:BHposMassGrowthMergers}. At this point, panels (a)-(b) reveal a transition from rotation-dominated to inflow-dominated gas dynamics and, concurrently, both the gas density $\rho$ and the number density increase, indicating the outbreak of enhanced accretion, as also showcased by panel (f). 

After $\sim 2$ Gyr, the accretion rate sharply decreases, following the decrease in $\rho$ and number density (panels d and e). At this moment, we observe massive spikes in local $\lvert v_{\rm bulk}\rvert$ (panel c). This increase in particle speeds also negatively impacts the accretion rate, as described by Eq.\eqref{eq:macc_rate}.  Lastly, the accretion rates of the central BH in both simulations follow different paths as the BH orbit approaches the center of the MW. The accretion rate increases for \code{M20BH}, while it decreases for \code{M80BH} at around $\gtrsim 4$ Gyr. However, at this point, the dwarf galaxy structure has been torn apart by the MW and the BH is left wandering, as shown in Figure~\ref{fig:BHposMassGrowthMergers}. This behavior is consistent with the fact that \code{M20BH} is expected to reach higher gas densities in the central region, since the supernova feedback is weaker, allowing the gas to cool more efficiently and accumulate in the inner regions.

\begin{figure*}
    \centering
    \includegraphics[width=0.8\linewidth]{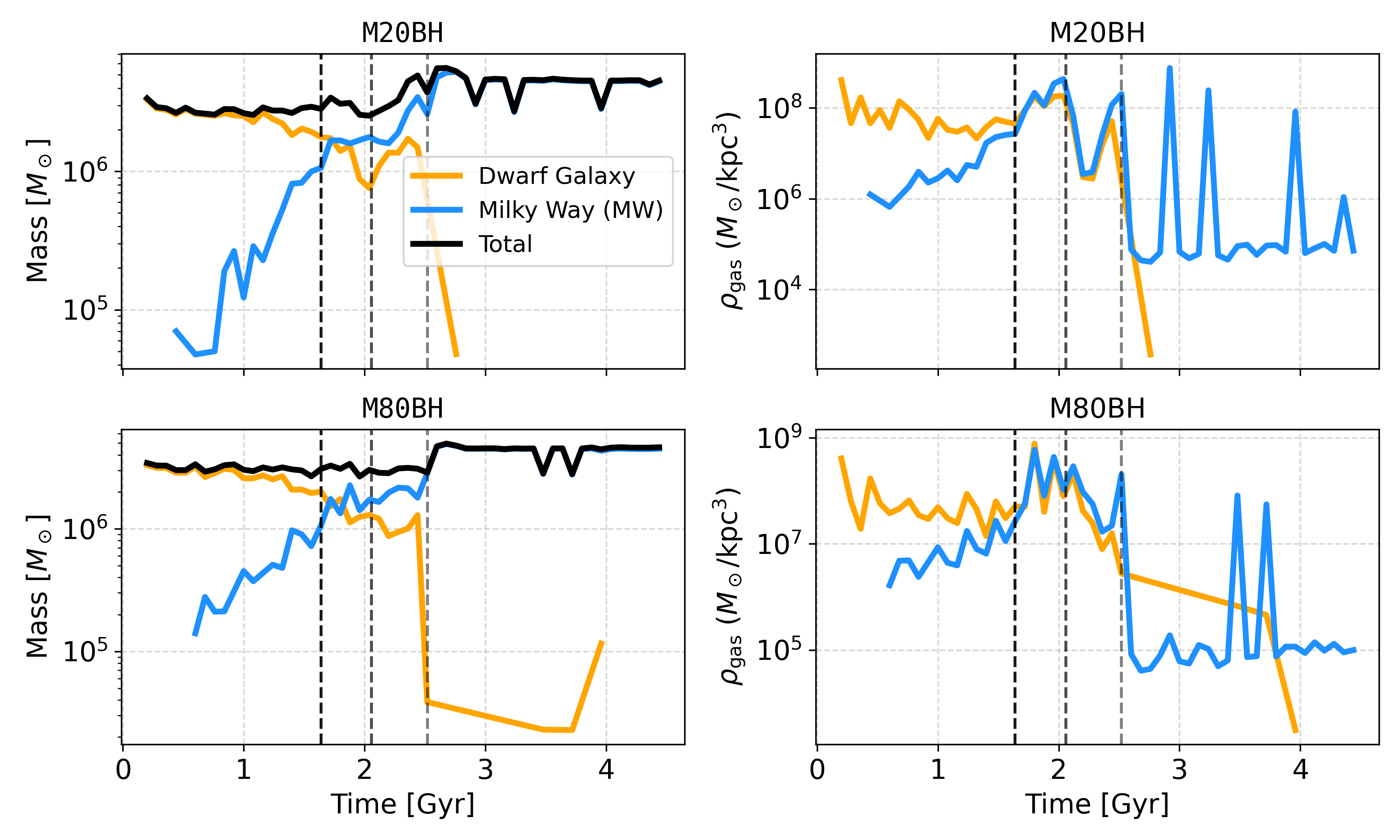}
    \caption{Time evolution of the gas mass and density in the vicinity of the BH, separated according to the initial galaxy of origin. The dashed lines denote the first pericenter, apocenter, and second pericenter.}
    \label{fig:gas_composition}
\end{figure*}

Furthermore, we track the origin of the 64 gas particles nearest to the BH throughout the simulations, as shown in Figure~\ref{fig:gas_composition}. In both \code{M20BH} and \code{M80BH}, there is an increasing contribution from gas originating in the MW-like galaxy as the dwarf approaches its first pericenter. Immediately after the first pericentric passage, the density of both gas components increases simultaneously, indicating a compression of the gas in the vicinity of the BH. This suggests that the enhanced accretion rate observed after the first pericentric passage is driven primarily by the compression of the surrounding gas associated with the deceleration of the dwarf galaxy, together with changes in the orbital motion of the gas around the BH.

\begin{figure*}
    \centering
    \includegraphics[width=\linewidth]{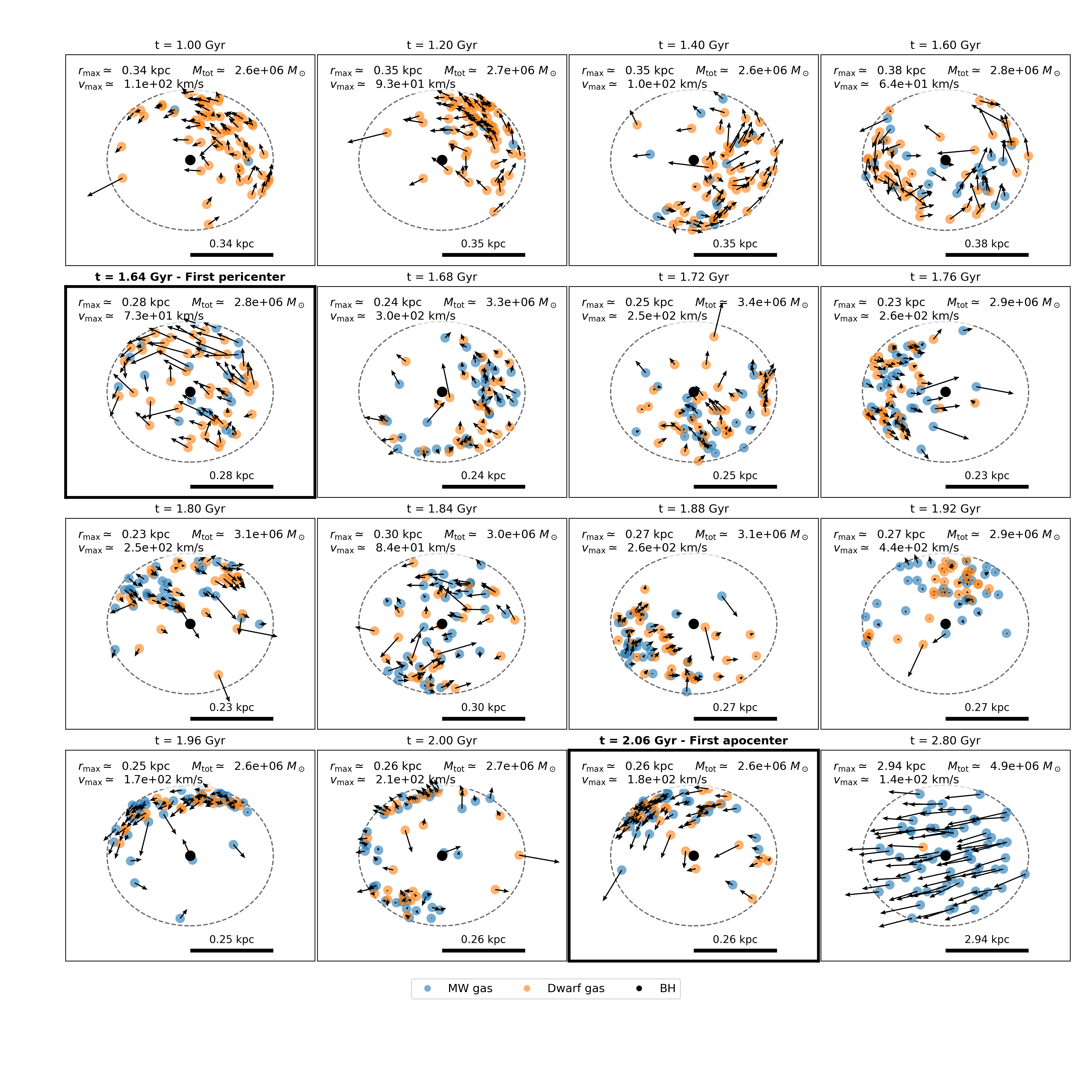}
    \caption{2D positions of the 64 nearest particles to the central BH across time for \code{M20BH}. Particles coming from the MW-like and the dwarf galaxy are represented by blue and orange circles, respectively. The arrows depict the velocity of those particles, and their sizes are normalized such that the highest velocity corresponds to $0.5\, r_{\rm max}$. We also show the total mass within the sphere defined by the longest particle's distance $r_{\rm max}$.}
    \label{fig:particles_near_BH_20}
\end{figure*}

\begin{figure*}
    \centering
    \includegraphics[width=\linewidth]{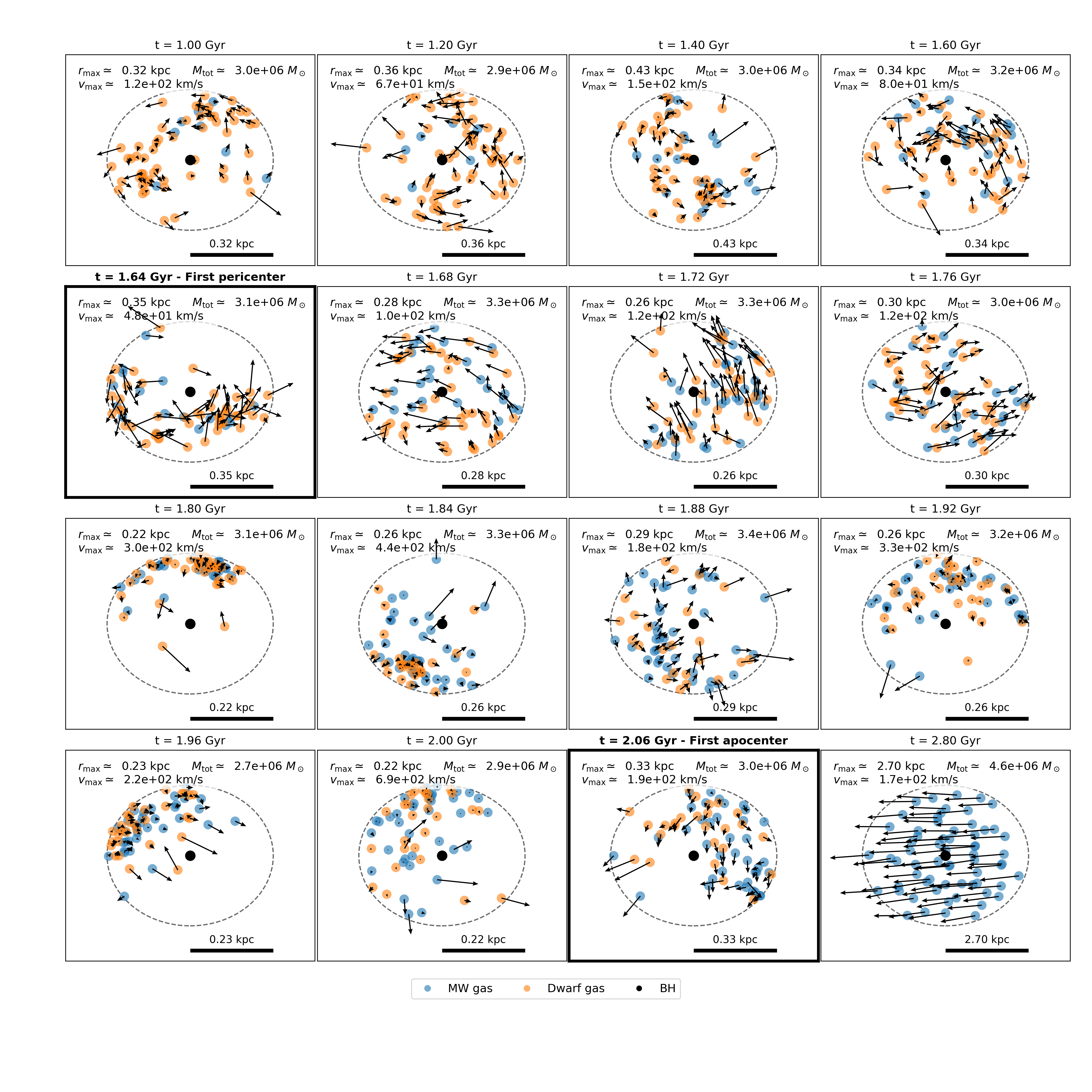}
    \caption{2D positions of the 64 nearest particles to the central BH across time for \code{M80BH}. Particles coming from the MW-like and the dwarf galaxy are represented by blue and orange circles, respectively. The arrows depict the velocity of those particles, and their sizes are normalized such that the highest velocity corresponds to $0.5\, r_{\rm max}$. We also show the total mass within the sphere defined by the longest particle's distance $r_{\rm max}$.}
    \label{fig:particles_near_BH_80}
\end{figure*}

This interpretation is further supported by the spatial distribution of the 64 nearest particles, shown for \code{M20BH} and \code{M80BH} in Figures~\ref{fig:particles_near_BH_20} and~\ref{fig:particles_near_BH_80}, respectively. Between $t = 1.60\,\rm{Gyr}$ and $t = 2.0\,\rm{Gyr}$, both simulations exhibit a reduction in the radius of the furthest of the 64 particles ($r_{\rm max}$) without a substantial change in the enclosed mass, indicating that the particles become more spatially concentrated. At the same time, the particles increasingly accumulate on one side of the circumference, providing further evidence of the compression and asymmetric redistribution of the gas around the BH.

We also observe pronounced decreases in the individual particle speeds during specific periods not necessarily linked to the pericenter and apocenter times. These reductions in velocity weaken the rotational support of the gas, which can make the inward transport of material easier, contributing to the enhanced accretion onto the BH. Finally, at $t=2.8$ Gyr (bottom right panel), the 64 nearest neighbors are composed solely of the host galaxy particles, which are scattered around a much larger volume, decreasing gas density and halting the accretion.

\subsection{Effects of the Merger with BH in the Dwarf Galaxy}

\begin{figure*}
    \centering
    \includegraphics[width=\linewidth]{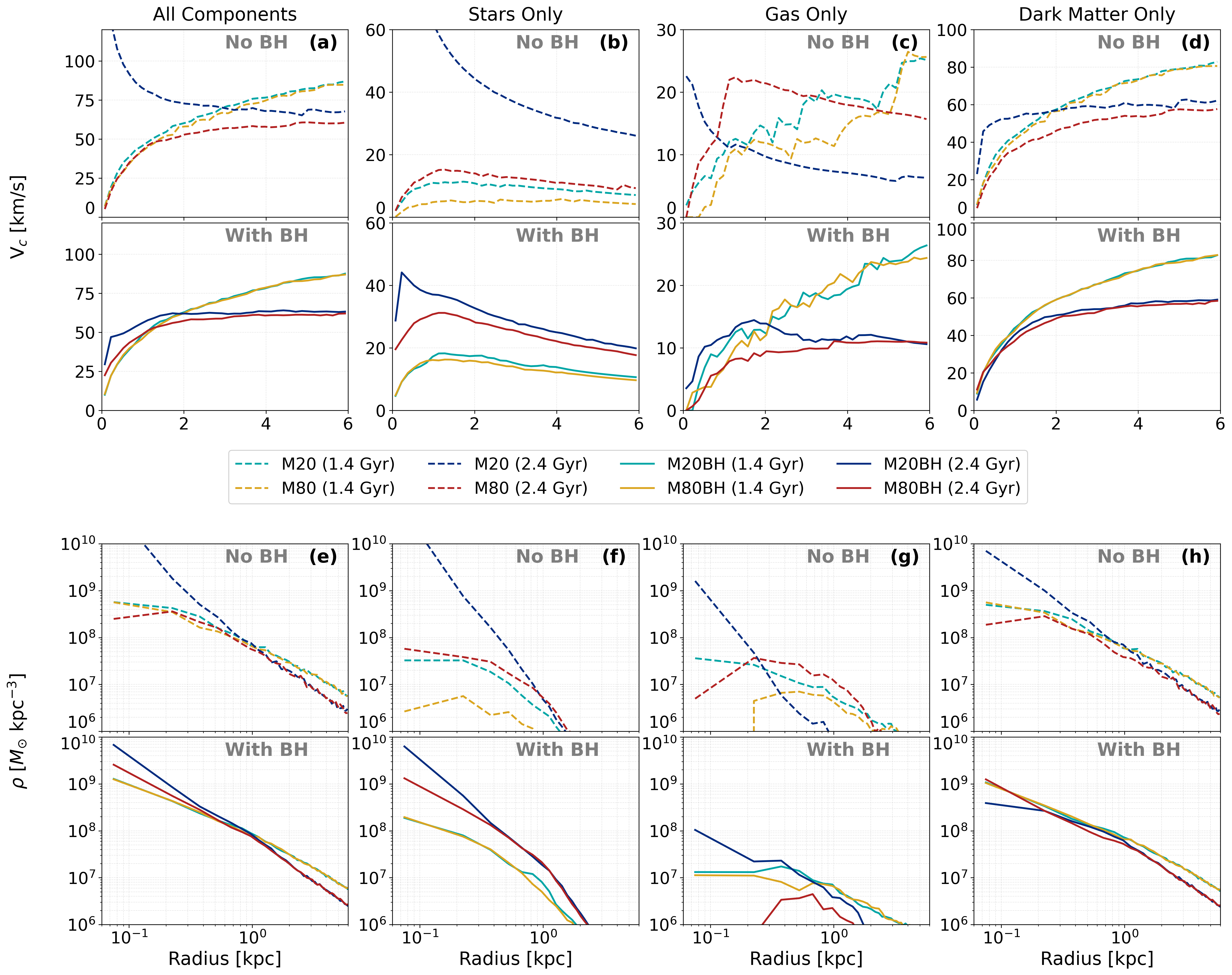}
    \caption{Evolution of the dwarf galaxy’s circular velocity profiles, $v_c$ (top row, panels a–d), and spherical mass density profiles, $\rho$ (bottom row, panels e–h). The columns show different mass components from left to right: all components, stars only, gas only, and dark matter only. Each panel is divided horizontally: the top sub-panel displays simulations without a central BH (\code{M20} and \code{M80}, shown with dashed lines), while the bottom sub-panel shows simulations with a BH included (\code{M20BH} and \code{M80BH}, shown with solid lines). Cool-colored lines represent models with 20\% feedback efficiency, while warm-colored lines represent models with 80\% feedback efficiency. Within each color family, cyan/turquoise-colored curves correspond to t=1.4 Gyr (before the first pericenter passage), whereas the dark blue/red curves show the profiles at t=2.4 Gyr (after the first pericenter passage).}
    \label{fig:merger_profiles}
\end{figure*}

To highlight the impact of the central BH accretion and feedback on the dwarf galaxy's structural properties due to the accretion-rate burst shown in Figures~\ref{fig:BHposMassGrowthMergers} and~\ref{fig:AccretionParametersEvolution} and the merger, we analyze the dwarf galaxy’s circular velocity profile, $v_c$, and mass density, $\rho$, in Figure~\ref{fig:merger_profiles}. Additionally, to disentangle the effects driven by the merger dynamics from those induced by BH accretion and feedback, we also present the corresponding $v_c$ and $\rho$ profiles from simulations without a central BH, taken from \cite{Amarante_2022} and denoted as \code{M20} and \code{M80}, which assume stellar feedback efficiencies of $20\%$ and $80\%$ of $10^{51}\,\mathrm{erg}$ per supernova, respectively. We show these profiles at two representative snapshots, $t = 1.4,\mathrm{Gyr}$ and $t = 2.4,\mathrm{Gyr}$, corresponding to epochs before the first pericentric passage and after it.

The impact of feedback strength in the absence of a BH is most evident at $t = 2.4$ Gyr, where the \code{M20} and \code{M80} models diverge significantly due to their differing capacities to regulate central mass accumulation. In the low-feedback \code{M20} model, the dwarf develops a massive central spike in gas density (panel g), which facilitates an enhanced star formation rate and results in a highly concentrated stellar distribution at $r \sim 0.1$ kpc (panel f). This centrally concentrated baryonic mass triggers the adiabatic contraction of the DM halo \citep{Blumenthal_1986, Gnedin_2004, Bullock_2017}, evidenced by a steeper and more elevated DM density profile at small radii compared to the higher-feedback \code{M80} counterpart (panel h). These structural transformations are mirrored in the rotation curves; while both models exhibit nearly identical behavior prior to the first pericentric passage at $t = 1.4$ Gyr, the \code{M20} stellar rotation curve at $t = 2.4$ Gyr develops a sharp central peak reaching $\sim 40$ km/s, whereas the \code{M80} profile remains significantly more extended and with a lower amplitude (panel b). When a central BH is included, it introduces a regulatory mechanism that moderates these feedback-driven extremes. In the low-feedback \code{M20BH} model, BH feedback and accretion suppress the massive gas spike and stellar concentration observed in the \code{M20} case (panels~f and~g), leading to a slightly less concentrated DM profile (panel h). Conversely, in the high-feedback regime, the \code{M80BH} simulation exhibits higher central gas and stellar densities than its \code{M80} counterpart, resulting in a more pronounced adiabatic contraction of the DM halo.

\begin{figure}
    \centering
    \includegraphics[width=\linewidth]{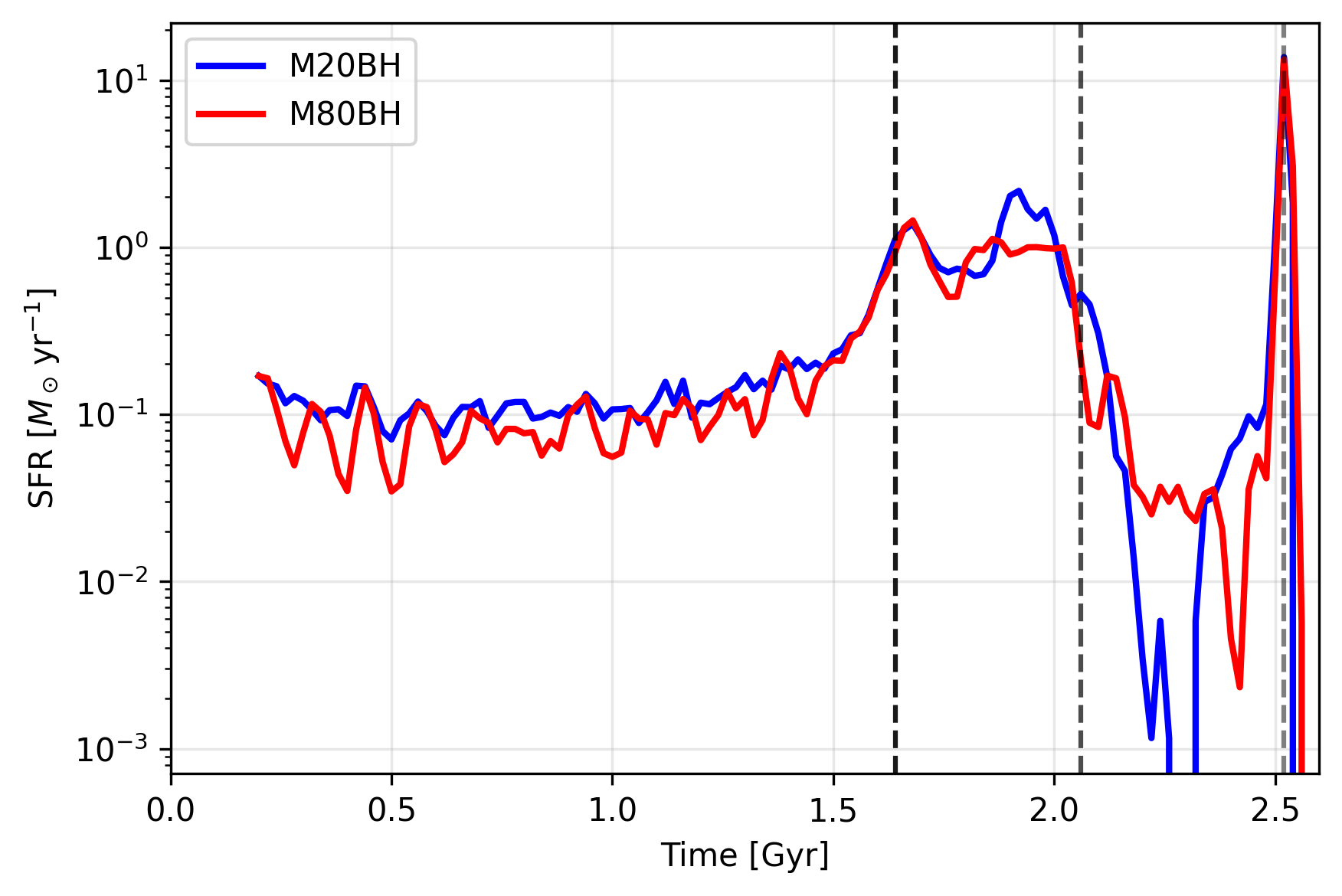}
    \caption{Evolution of the SFR of the dwarf galaxy over $2.6$ Gyr for \code{M20BH} (blue) and \code{M80BH} (red) models. The vertical dashed lines mark the times of the first pericenter, the first apocenter, and the second pericenter.}
    \label{fig:SFR_dwarf}
\end{figure}

To further elucidate the physical processes observed in the dwarf galaxy, we display in Figure~\ref{fig:SFR_dwarf} the time evolution of its SFR for the \code{M20BH} and \code{M80BH} models. In both cases, the SFR exhibits an initial quasi-steady phase, followed by a pronounced tidally induced starburst coincident with the first pericentric passage, previously reported in \cite{Amarante_2022}. Interestingly, this epoch also coincides with the most metal-rich GSE globular cluster ages inferred from relative ages of GSE and in-situ ones from the GSE-induced MW starburst \citep[see figure 3 of][]{laporte2026ocen}, providing observational support for the tidally induced starburst predicted in our models. This burst is rapidly followed by a strong suppression of star formation in both models, although the effect is significantly more severe in \code{M20BH}, which undergoes a complete quenching of star formation at $t \simeq 2.1\,\mathrm{Gyr}$. In contrast, the \code{M80BH} model sustains a relatively steady SFR of $\sim 0.05\,{M_\odot\,{\rm yr}^{-1}}$ over the same period. At later times, around $t \simeq 2.7\,\mathrm{Gyr}$, the SFR in both simulations rises sharply to ${\sim}10\,{M_\odot\,{\rm yr}^{-1}}$, marking the final coalescence of the dwarf galaxy with the MW as seen in panels (a) and (b) of Figure~\ref{fig:BHposMassGrowthMergers}. 

The stronger quenching observed in the \code{M20BH} model appears to be closely linked to its enhanced BH accretion history. As shown in panels (e) and (f) of Figure~\ref{fig:BHposMassGrowthMergers}, the BH in \code{M20BH} experiences greater growth in mass when compared to \code{M80BH}. This difference suggests that a larger fraction of the available gas in \code{M20BH} is funneled toward the central BH during the galaxy evolution, reducing the amount of cold gas available for star formation. As a result, star formation is suppressed in \code{M20BH}, whereas the slightly weaker accretion in \code{M80BH} allows star formation to persist at a low but non-zero level.

\section{Discussion}
\label{sec:discussions}
One of the key findings in this paper is that the merger between a GSE-type dwarf galaxy and a MW–like galaxy can trigger enhanced gas inflows that significantly increase BH accretion and alter the evolution of the BH. This behavior is well established in massive galaxies \citep{Hopkins_2006, Wild_2007}. However, recent findings also show a link between an increased fraction of AGNs in dwarf galaxies and mergers \citep{Er_stegui_2025}, where the first close encounter also enhances BH accretion the most.
From cosmological simulations, the BH occupation fraction in dwarf galaxies is relatively high, but the AGN fraction remains very low, indicating that most BHs are quiescent \citep{Bellovary_2019}. This reflects the fact that sustained accretion episodes are rare and require favorable conditions such as central gas concentration and dynamical perturbations.

In our isolated runs, supernova feedback plays a more dominant role than the central BH in regulating the early evolution of the galaxy. This is consistent with cosmological simulations by \citet{Dubois_2015} and \citet{Habouzit_2017}, which showed that supernova feedback regulates the availability of cold gas in the galactic nucleus and consequently controls early BH growth. In particular, \citet{Dubois_2015} found that weaker supernova feedback allows a larger central gas reservoir to persist, leading to rapid BH growth followed by self-regulation. This behavior resembles the rapid initial growth and subsequent stabilization observed in \code{M20BH}, as shown in Figure~\ref{fig:IsoBHGrowthAccRateRhoRadialProf}(a). Furthermore, as the central mass concentration increases, the galaxy becomes more efficient at retaining supernova-driven outflows, which can contribute to the enhanced accretion rates observed at later times in our \code{M80BH} model.

In the interacting models, after the first pericentric passage, the central BH achieves an accretion rate of ${\sim} 0.25\,\dot{M}_{\rm Edd}$ for a period of ${\lesssim} 0.75$\,Gyr. This places the system within a radiatively efficient AGN regime \citep{Shakura_1973}, but the activity is clearly episodic rather than sustained over cosmological timescales. Comparable accretion rates have been observed in different mass environments, including late-stage dwarf-dwarf mergers \citep{Kimbrell_2024}, early-stage minor mergers where the dwarf companion undergoes a brief, intense phase of near-Eddington accretion \citep{Micic_2024}, and a heavily obscured low-mass AGN that is likely triggered by a galaxy merger \citep{Wang_2026}. 

In this spirit, \cite{Micic_2026} likewise reported a ${\sim}10{:}1$ minor merger hosting a low-mass black hole accreting at ${\sim}0.2$--$0.5,\dot M_{\rm Edd}$, providing observational support for episodic rapid growth during minor mergers. \cite{Capelo_2015} systematically simulated galaxy mergers across multiple mass ratios, including 1:10 minor mergers, also showing a similar, strictly episodic, burst to the central BH in minor mergers, with the high-accretion growth phase heavily constrained by the orbital timeline with bursts of order ${\sim}0.3$\,Gyr. At substantially higher mass resolution, \cite{Partmann_2025} found that nuclear star clusters can promote BH growth in isolated dwarf galaxies by funneling gas into the central few parsecs, while stellar feedback regulates accretion through episodic suppression of the nuclear gas supply. Unlike our simulations, however, their models do not include BH feedback. In addition to the episodic nature of the accretion enhancement, the disruption of the gas disk induced by the merger (evidenced by panels (a) and (b) of Figure~\ref{fig:AccretionParametersEvolution}) has been shown to enhance both accretion and feedback in cosmological simulations \citep{Pontzen_2017}. We note, however, that the coupling efficiency between BH feedback and the surrounding gas, represented by $\epsilon_{\rm f}$ in Eq.~\ref{eq:accretion_energy}, adopted in this work, is at the lower end of the range explored by \citet{Tremmel_2017}. This value was chosen empirically to ensure sustained growth in our isolated simulations, but remains lower than the values considered by \cite{Pontzen_2017}, which do not find a strong, sustained increase in accretion.

The final BH mass in our simulations exceeds that of the supermassive BH at the Galactic center, Sgr~A*, whose mass is well-constrained to $M \simeq 4 \times 10^{6} M_{\odot}$ \citep{Do_2019, Grav_colab_2019}, and the parameters of accretion have been extensively studied by the Event Horizon Telescope Collaboration \citep{EHTC_2022_V}. We emphasize that this large final mass is primarily determined by the large initial seed mass adopted in our initial conditions---chosen for computational reasons---rather than by nonphysical over-accretion driven by the merger dynamics. The BH growth in the early stages of the merger runs---while the dwarf galaxy is still far from the effects of the MW---is similar to that encountered in our higher-resolution isolated runs, which used a BH seed of $8 \times 10^3\, M_\odot$. Consequently, adopting a lower, more observationally motivated initial seed mass would systematically downscale the final BH mass, potentially to values below Sgr A*, while preserving the high-accretion burst presented in this work. 

As a rough estimation, assuming the same accretion rate, an initial seed value of $M_{\rm BH} = 100\,M_\odot$, more consistent with a stellar-origin seed than direct collapse \citep[e.g.,][]{Volonteri_2008, Volonteri_2010}, evolved under the same accretion history measured in our simulations, would grow to a final mass of $\sim 2.6\times10^{4}\,M_\odot$ in the \code{M20BH} model and $\sim 1.1\times10^{4}\,M_\odot$ in the \code{M80BH} model. This estimated final mass is comparable to recent dynamical mass estimates for the \ocen IMBH \citep{Haberle_2024}, and more than two orders of magnitude below Sgr~A*.

We highlight that, in the standard Bondi--Hoyle prescription, the accretion rate scales as $\dot{M}_{\rm BH}\propto M_{\rm BH}^2$, so reducing the initial black-hole mass while keeping the gas properties fixed would result in a substantially different accretion history. In our simulations, however, the Bondi--Hoyle accretion prescription is enhanced by a term that depends on $n^2$. When reducing the mass of the gas particles, and correspondingly the black-hole mass, we increase the particle number density proportionally in order to preserve the same black-hole-to-gas mass ratio during the merger. This compensates for the explicit dependence on $M_{\rm BH}$ and motivates using the same measured accretion history for this back-of-the-envelope estimate.

Finally, in our simulations, the presence of the central BH regulates the efficiency of supernova feedback in reshaping the inner DM distribution. In particular, for model \code{M20BH}, the BH–driven gas depletion reduces the central gas reservoir, which in turn suppresses subsequent star formation in the innermost regions. This leads to a weaker baryonic gravitational potential and therefore a diminished degree of adiabatic contraction of the DM halo. By contrast, in the \code{MBH80} model, the higher retained central baryonic mass sustains more concentrated star formation, resulting in a deeper central potential well and a correspondingly stronger contraction of the inner DM component. We find a different behavior in the corresponding simulations without a central BH (\code{M20} and \code{M80}; \citealt{Amarante_2022}), where the absence of BH-driven gas depletion allows for more sustained central star formation and for model \code{M20}, consequently, a more centrally concentrated baryonic potential, leading to a stronger response of the inner DM distribution.

\section{Conclusions}
\label{sec:conclusion}

In this work, we have performed high-resolution $N$-body + SPH numerical simulations using the \code{ChaNGa} framework to investigate the co-evolution and merger dynamics of a MW-like host galaxy and a colliding dwarf galaxy, mimicking a Gaia-Sausage-Enceladus (GSE) progenitor setup. Our models self-consistently incorporate metallicity-dependent cooling, turbulent gas diffusion, blast-wave supernova feedback ($\text{FB20}$ and $\text{FB80}$ regimes), and a subgrid approach for unresolved BH dynamical friction. Our setup features a central intermediate-mass BH ($M_{\rm BH} = 8 \times 10^3 M_\odot$) in the isolated runs and $M_{\rm BH} = 10^{5} M_{\odot}$ in the merger runs, seeded within the dwarf galaxy, governed by a modified, angular-momentum-aware Bondi-Hoyle accretion prescription.

Although we have utilized a modified accretion model that accounts for rotational support, it is important to acknowledge the caveats of this approach. The modified Bondi-Hoyle prescription remains a subgrid approximation that is unable to fully capture the small-scale physics of the gas feeding the central BH. In reality, General Relativistic Magnetohydrodynamic (GRMHD) models provide a far more accurate description of accretion flows, magnetic fields, and angular momentum transport near the event horizon. Although some current works are tackling this problem \citep[e.g.,][]{Cho_2026}, bridging the immense dynamic range required to self-consistently couple these horizon-scale GRMHD simulations with kpc-scale galaxy mergers remains a challenge. Furthermore, the accretion rate measured in this work, at times, is expected to lie in the radiatively efficient regime, where radiation processes will play a significant role in regulating both the structure of the accretion flow and the associated feedback energetics, a regime typically treated within radiative GRMHD frameworks, which self-consistently capture the coupled dynamics of radiation, magnetic fields, and relativistic flows near the event horizon.

Consequently, many physical mechanisms are omitted in our current setup. All quantities governing the active galactic nucleus are simplified; for instance, the BH radiative efficiency ($\eta_{\rm rad}$) is kept fixed, and explicit kinetic feedback channels—such as the collimation of relativistic jets or the driving of powerful, large-scale winds—are non-existent in our model. Future work will need to explore more sophisticated accretion and feedback frameworks that can better bridge these scales, capturing the interplay between small-scale kinetic outflows and large-scale galactic evolution.
From the simulations performed in this work, we highlight the following results:
\begin{enumerate}
    \item The simulations depicting an isolated dwarf galaxy demonstrate that without a merger event to drive gas accretion, the BH undergoes very little growth (${\sim} 3$ times its seed mass) over a period of 5\,Gyr, with a low accretion rate. This difficulty is consistent with findings reported in other recent works \citep{Hazenfratz2025dwarfBH, Petersson_2025}. 
    \item The simulations including the host MW-like galaxy merger demonstrate that the central BH accretion is not quenched by tidal and/or ram-pressure stripping of the gas, as originally conjectured by \citet{Limberg_2024}. Instead, the gas around the central BH compresses and gets slowly fueled by the MW-like's gas reservoir, accelerating its growth after the first pericentric approach. This compression drives massive mass growth in both models, resulting in a two orders of magnitude increase in the initial BH mass by the time the dwarf is fully disrupted.
    \item The burst in accretion rate experienced by the BH during the first pericentric passage plays a significant role in regulating and suppressing star formation within the dwarf galaxy's central regions. Following a tidally induced starburst at the first pericentric passage, the subsequent injection of BH feedback energy drives a strong suppression of the SFR in both merger models. This effect is particularly severe in the \code{M20BH} model, where the lower supernova feedback allows for more extreme BH growth, fueling a stronger AGN feedback to completely quench star formation after the first pericentric approach at $t \simeq 2.1\ \rm{Gyr}$. In contrast, the stronger stellar feedback in \code{M80BH} maintains a more diffuse gas reservoir, allowing the dwarf to sustain a low-level, steady SFR until the final coalescence at $t \simeq 2.7\,\rm{Gyr}$.
    \item The presence of the BH in the simulations of the isolated dwarf galaxy (\code{ISOBH20} and \code{ISOBH80}) seems to have a negligible impact on the structural properties of the dwarf galaxy when compared to simulations without the BH presence (\code{ISO20} and \code{ISO80}). The simulations confirm that, as expected for low-mass systems, supernova feedback is the primary mechanism regulating the physical state of the interstellar medium and controlling star formation, hence chemical enrichment, within the dwarf galaxy.

    \item The presence of the central BH affects the inner DM response by modulating the central baryonic potential. In the merger runs, BH-driven gas depletion in \code{M20BH} reduces central star formation and weakens the baryonic potential, leading to a reduced degree of adiabatic contraction. In contrast, \code{M80BH} retains a more concentrated central baryonic component, resulting in a deeper potential well and a stronger contraction of the inner DM halo. In the corresponding runs without a central BH (\code{M20} and \code{M80}; \citealt{Amarante_2022}), sustained central star formation leads to a more concentrated baryonic potential, enhancing the response of the DM distribution, particularly in \code{M20}.
\end{enumerate}

The results presented in this work are by no means exhaustive. Despite using a more advanced, angular-momentum-aware accretion model that improves upon standard Bondi-Hoyle prescriptions, significant small-scale and large-scale physics remain missing from the galactic modeling components of our simulations. Nevertheless, numerical experiments such as the ones performed here will be crucial to interpret not only the \ocen-GSE connection in the Milky Way, but also current and future observations of both overmassive central BHs and AGNs in dwarf galaxies.

\begin{acknowledgments}
We thank Michael Tremmel for insightful comments on the manuscript.
P.N.M. acknowledges financial support from the Fundação de Amparo à Pesquisa do Estado de São Paulo (FAPESP) under grant number 2023/15835-2 and 2026/02386-3.
J.A.S.A. and Z.Y.L. are supported by the National Natural Science Foundation of China under grant Nos. 12233001, 12533004, by the National Key R\&D Program of China under grant No. 2024YFA1611602, by a Shanghai Natural Science Research Grant (24ZR1491200), by the ``111'' project of the Ministry of Education under grant No. B20019, and by the China Manned Space Program with grant Nos. CMS-CSST-2025-A08, CMS-CSST-2025-A09 and CMS-CSST-2025-A11. 
L.B. thanks the financial support by the São Paulo Research Foundation (FAPESP), Brasil (Proc. 2024/16510-2). 
C.L. acknowledges funding from the European Research Council (ERC) under the European Union’s Horizon 2020 research and innovation programme (grant agreement No. 852839) and the Agence Nationale de la Recherche (ANR project ANR-24-CPJ1-0160-01).
R.S.L. is supported by the Fundação de Amparo à Pesquisa do Estado de São Paulo (FAPESP), Process Number 2025/11756-6. 
This work has used resources from the MareNostrum 4 supercomputer at the Barcelona Supercomputing Center (BSC). Further analysis and simulation work were performed on the Gravity Supercomputer at the Department of Astronomy, Shanghai Jiao Tong University, and on the NYX supercomputer at the Universitat de Barcelona (ICCUB).
\end{acknowledgments}

\bibliography{citations}{}

@ARTICLE{Volonteri_2008,
       author = {{Volonteri}, Marta and {Lodato}, Giuseppe and {Natarajan}, Priyamvada},
        title = "{The evolution of massive black hole seeds}",
      journal = {\mnras},
         year = 2008,
        month = jan,
       volume = {383},
       number = {3},
        pages = {1079-1088},
          doi = {10.1111/j.1365-2966.2007.12589.x},
archivePrefix = {arXiv},
       eprint = {0709.0529},
 primaryClass = {astro-ph},
       adsurl = {https://ui.adsabs.harvard.edu/abs/2008MNRAS.383.1079V}
}

@ARTICLE{tomlinson+2026,
       author = {{Tomlinson}, Thomas and {Fragkoudi}, Francesca and {Carrillo}, Andreia and {Fattahi}, Azadeh and {Gherghinescu}, Paula and {Deason}, Alis and {Pakmor}, R{\"u}diger and {Grand}, Robert J.~J. and {G{\'o}mez}, Facundo A. and {van de Voort}, Freeke and {Bieri}, Rebekka},
        title = "{Stirring things up: Bar-induced substructures in the stellar halo of a cosmological Milky Way analogue}",
      journal = {\mnras},
         year = 2026,
        month = jun,
       volume = {549},
       number = {1},
          eid = {stag841},
        pages = {stag841},
          doi = {10.1093/mnras/stag841},
archivePrefix = {arXiv},
       eprint = {2601.14409},
 primaryClass = {astro-ph.GA},
       adsurl = {https://ui.adsabs.harvard.edu/abs/2026MNRAS.549ag841T}
}

@ARTICLE{deg+2019,
       author = {{Deg}, N. and {Widrow}, L.~M. and {Randriamampandry}, T. and {Carignan}, C.},
        title = "{GALACTICS with gas}",
      journal = {\mnras},
         year = 2019,
        month = jul,
       volume = {486},
       number = {4},
        pages = {5391-5399},
          doi = {10.1093/mnras/stz1203},
archivePrefix = {arXiv},
       eprint = {1904.12700},
 primaryClass = {astro-ph.GA},
       adsurl = {https://ui.adsabs.harvard.edu/abs/2019MNRAS.486.5391D}
}

@article{Governato_1994,
   title={The fate of central black holes in merging galaxies},
   volume={271},
   ISSN={1365-2966},
   url={http://dx.doi.org/10.1093/mnras/271.2.317},
   DOI={10.1093/mnras/271.2.317},
   number={2},
   journal={Monthly Notices of the Royal Astronomical Society},
   publisher={Oxford University Press (OUP)},
   author={Governato, F. and Colpi, M. and Maraschi, L.},
   year={1994},
   month=Nov, pages={317–322} }

@ARTICLE{McWilliam_1997,
       author = {{McWilliam}, Andrew},
        title = "{Abundance Ratios and Galactic Chemical Evolution}",
      journal = {\araa},
         year = 1997,
        month = jan,
       volume = {35},
        pages = {503-556},
          doi = {10.1146/annurev.astro.35.1.503},
       adsurl = {https://ui.adsabs.harvard.edu/abs/1997ARA&A..35..503M}
}

@ARTICLE{DEFW_1985,
       author = {{Davis}, M. and {Efstathiou}, G. and {Frenk}, C.~S. and {White}, S.~D.~M.},
        title = "{The evolution of large-scale structure in a universe dominated by cold dark matter}",
      journal = {\apj},
         year = 1985,
        month = may,
       volume = {292},
        pages = {371-394},
          doi = {10.1086/163168},
       adsurl = {https://ui.adsabs.harvard.edu/abs/1985ApJ...292..371D}
}

@article{Croton_2006,
   title={The many lives of active galactic nuclei: cooling flows, black holes and the luminosities and colours of galaxies},
   volume={365},
   ISSN={1365-2966},
   url={http://dx.doi.org/10.1111/j.1365-2966.2005.09675.x},
   DOI={10.1111/j.1365-2966.2005.09675.x},
   number={1},
   journal={Monthly Notices of the Royal Astronomical Society},
   publisher={Oxford University Press (OUP)},
   author={Croton, Darren J. and Springel, Volker and White, Simon D. M. and De Lucia, G. and Frenk, C. S. and Gao, L. and Jenkins, A. and Kauffmann, G. and Navarro, J. F. and Yoshida, N.},
   year={2006},
   month=Jan, pages={11–28} }

@ARTICLE{Kauffmann_2000,
       author = {{Kauffmann}, Guinevere and {Haehnelt}, Martin},
        title = "{A unified model for the evolution of galaxies and quasars}",
      journal = {\mnras},
         year = 2000,
        month = jan,
       volume = {311},
       number = {3},
        pages = {576-588},
          doi = {10.1046/j.1365-8711.2000.03077.x},
archivePrefix = {arXiv},
       eprint = {astro-ph/9906493},
 primaryClass = {astro-ph},
       adsurl = {https://ui.adsabs.harvard.edu/abs/2000MNRAS.311..576K}
}

@ARTICLE{Lynden_1969,
       author = {{Lynden-Bell}, D.},
        title = "{Galactic Nuclei as Collapsed Old Quasars}",
      journal = {\nat},
         year = 1969,
        month = aug,
       volume = {223},
       number = {5207},
        pages = {690-694},
          doi = {10.1038/223690a0},
       adsurl = {https://ui.adsabs.harvard.edu/abs/1969Natur.223..690L}
}

@ARTICLE{Myeong_2019,
       author = {{Myeong}, G.~C. and {Vasiliev}, E. and {Iorio}, G. and {Evans}, N.~W. and {Belokurov}, V.},
        title = "{Evidence for two early accretion events that built the Milky Way stellar halo}",
      journal = {\mnras},
         year = 2019,
        month = sep,
       volume = {488},
       number = {1},
        pages = {1235-1247},
          doi = {10.1093/mnras/stz1770},
archivePrefix = {arXiv},
       eprint = {1904.03185},
 primaryClass = {astro-ph.GA},
       adsurl = {https://ui.adsabs.harvard.edu/abs/2019MNRAS.488.1235M}
}

@article{Wild_2007,
   title={Bursty stellar populations and obscured active galactic nuclei in galaxy bulges},
   volume={381},
   ISSN={1365-2966},
   url={http://dx.doi.org/10.1111/j.1365-2966.2007.12256.x},
   DOI={10.1111/j.1365-2966.2007.12256.x},
   number={2},
   journal={Monthly Notices of the Royal Astronomical Society},
   publisher={Oxford University Press (OUP)},
   author={Wild, V. and Kauffmann, G. and Heckman, T. and Charlot, S. and Lemson, G. and Brinchmann, J. and Reichard, T. and Pasquali, A.},
   year={2007},
   month=Oct, pages={543–572} }

@ARTICLE{Hopkins_2006,
       author = {{Hopkins}, Philip F. and {Hernquist}, Lars and {Cox}, Thomas J. and {Di Matteo}, Tiziana and {Robertson}, Brant and {Springel}, Volker},
        title = "{A Unified, Merger-driven Model of the Origin of Starbursts, Quasars, the Cosmic X-Ray Background, Supermassive Black Holes, and Galaxy Spheroids}",
      journal = {\apjs},
         year = 2006,
        month = mar,
       volume = {163},
       number = {1},
        pages = {1-49},
          doi = {10.1086/499298},
archivePrefix = {arXiv},
       eprint = {astro-ph/0506398},
 primaryClass = {astro-ph},
       adsurl = {https://ui.adsabs.harvard.edu/abs/2006ApJS..163....1H}
}

@ARTICLE{Grav_colab_2019,
       author = {{GRAVITY Collaboration} and {Abuter}, R. and {Amorim}, A. and {Baub{\"o}ck}, M. and {Berger}, J.~P. and {Bonnet}, H. and {Brandner}, W. and {Cl{\'e}net}, Y. and {Coud{\'e} Du Foresto}, V. and {de Zeeuw}, P.~T. and {Dexter}, J. and {Duvert}, G. and {Eckart}, A. and {Eisenhauer}, F. and {F{\"o}rster Schreiber}, N.~M. and {Garcia}, P. and {Gao}, F. and {Gendron}, E. and {Genzel}, R. and {Gerhard}, O. and {Gillessen}, S. and {Habibi}, M. and {Haubois}, X. and {Henning}, T. and {Hippler}, S. and {Horrobin}, M. and {Jim{\'e}nez-Rosales}, A. and {Jocou}, L. and {Kervella}, P. and {Lacour}, S. and {Lapeyr{\`e}re}, V. and {Le Bouquin}, J.-B. and {L{\'e}na}, P. and {Ott}, T. and {Paumard}, T. and {Perraut}, K. and {Perrin}, G. and {Pfuhl}, O. and {Rabien}, S. and {Rodriguez Coira}, G. and {Rousset}, G. and {Scheithauer}, S. and {Sternberg}, A. and {Straub}, O. and {Straubmeier}, C. and {Sturm}, E. and {Tacconi}, L.~J. and {Vincent}, F. and {von Fellenberg}, S. and {Waisberg}, I. and {Widmann}, F. and {Wieprecht}, E. and {Wiezorrek}, E. and {Woillez}, J. and {Yazici}, S.},
        title = "{A geometric distance measurement to the Galactic center black hole with 0.3\% uncertainty}",
      journal = {\aap},
         year = 2019,
        month = may,
       volume = {625},
          eid = {L10},
        pages = {L10},
          doi = {10.1051/0004-6361/201935656},
archivePrefix = {arXiv},
       eprint = {1904.05721},
 primaryClass = {astro-ph.GA},
       adsurl = {https://ui.adsabs.harvard.edu/abs/2019A&A...625L..10G}
}

@article{Capelo_2015,
   title={Growth and activity of black holes in galaxy mergers with varying mass ratios},
   volume={447},
   ISSN={0035-8711},
   url={http://dx.doi.org/10.1093/mnras/stu2500},
   DOI={10.1093/mnras/stu2500},
   number={3},
   journal={Monthly Notices of the Royal Astronomical Society},
   publisher={Oxford University Press (OUP)},
   author={Capelo, Pedro R. and Volonteri, Marta and Dotti, Massimo and Bellovary, Jillian M. and Mayer, Lucio and Governato, Fabio},
   year={2015},
   month=Jan, pages={2123–2143} }

@ARTICLE{Do_2019,
       author = {{Do}, Tuan and {Hees}, Aurelien and {Ghez}, Andrea and {Martinez}, Gregory D. and {Chu}, Devin S. and {Jia}, Siyao and {Sakai}, Shoko and {Lu}, Jessica R. and {Gautam}, Abhimat K. and {O'Neil}, Kelly Kosmo and {Becklin}, Eric E. and {Morris}, Mark R. and {Matthews}, Keith and {Nishiyama}, Shogo and {Campbell}, Randy and {Chappell}, Samantha and {Chen}, Zhuo and {Ciurlo}, Anna and {Dehghanfar}, Arezu and {Gallego-Cano}, Eulalia and {Kerzendorf}, Wolfgang E. and {Lyke}, James E. and {Naoz}, Smadar and {Saida}, Hiromi and {Sch{\"o}del}, Rainer and {Takahashi}, Masaaki and {Takamori}, Yohsuke and {Witzel}, Gunther and {Wizinowich}, Peter},
        title = "{Relativistic redshift of the star S0-2 orbiting the Galactic Center supermassive black hole}",
      journal = {Science},
         year = 2019,
        month = aug,
       volume = {365},
       number = {6454},
        pages = {664-668},
          doi = {10.1126/science.aav8137},
archivePrefix = {arXiv},
       eprint = {1907.10731},
 primaryClass = {astro-ph.GA},
       adsurl = {https://ui.adsabs.harvard.edu/abs/2019Sci...365..664D}
}

@article{Wang_2026,
doi = {10.3847/1538-4357/ae6779},
url = {https://doi.org/10.3847/1538-4357/ae6779},
year = {2026},
month = {may},
publisher = {The American Astronomical Society},
volume = {1003},
number = {2},
pages = {172},
author = {Wang, Shouyi and Zou, Fan and Chen, Chang-Hao and Brandt, W. N. and Gallo, Elena and Luo, Bin and Wu, Xue-Bing and Fu, Yuming and Nguyen, Dieu D. and Sun, Shengxiu},
title = {A Rare Eddington-limited, Heavily Obscured Low-mass Active Galactic Nucleus Likely Triggered by a Galaxy Merger},
journal = {The Astrophysical Journal}
}

@article{Micic_2024,
   title={SDSS J125417.98+274004.6: An X-ray Detected Minor Merger Dual AGN},
   volume={7},
   ISSN={2565-6120},
   url={http://dx.doi.org/10.21105/astro.2310.09945},
   DOI={10.21105/astro.2310.09945},
   journal={The Open Journal of Astrophysics},
   publisher={Maynooth University},
   author={Mićić, Marko and Wells, Brenna N. and Holmes, Olivia J. and Irwin, Jimmy A.},
   year={2024},
   month=Jan }

@article{Kimbrell_2024,
doi = {10.3847/1538-4357/ad712d},
url = {https://doi.org/10.3847/1538-4357/ad712d},
year = {2024},
month = {oct},
publisher = {The American Astronomical Society},
volume = {974},
number = {1},
pages = {51},
author = {Kimbrell, Seth J. and Reines, Amy E.},
title = {A Luminous X-Ray Active Galactic Nucleus in the Dwarf–Dwarf Galaxy Merger RGG 66},
journal = {The Astrophysical Journal}
}

@article{Bellovary_2019,
    author = {Bellovary, Jillian M and Cleary, Colleen E and Munshi, Ferah and Tremmel, Michael and Christensen, Charlotte R and Brooks, Alyson and Quinn, Thomas R},
    title = {Multimessenger signatures of massive black holes in dwarf galaxies},
    journal = {Monthly Notices of the Royal Astronomical Society},
    volume = {482},
    number = {3},
    pages = {2913-2923},
    year = {2019},
    month = {01},
    issn = {0035-8711},
    doi = {10.1093/mnras/sty2842},
    url = {https://doi.org/10.1093/mnras/sty2842},
    eprint = {https://academic.oup.com/mnras/article-pdf/482/3/2913/26607934/sty2842.pdf},
}

@article{Er_stegui_2025,
   title={MaNGA AGN dwarf galaxies (MAD): III. The role of mergers and environment in active galactic nucleus activity in dwarf galaxies},
   volume={699},
   ISSN={1432-0746},
   url={http://dx.doi.org/10.1051/0004-6361/202554387},
   DOI={10.1051/0004-6361/202554387},
   journal={Astronomy \&; Astrophysics},
   publisher={EDP Sciences},
   author={Eróstegui, A. and Mezcua, M. and Siudek, M. and Domínguez Sánchez, H. and Rodríguez Morales, V.},
   year={2025},
   month=July, pages={A330} }

@ARTICLE{Bullock_2017,
       author = {{Bullock}, James S. and {Boylan-Kolchin}, Michael},
        title = "{Small-Scale Challenges to the {\ensuremath{\Lambda}}CDM Paradigm}",
      journal = {\araa},
         year = 2017,
        month = aug,
       volume = {55},
       number = {1},
        pages = {343-387},
          doi = {10.1146/annurev-astro-091916-055313},
archivePrefix = {arXiv},
       eprint = {1707.04256},
 primaryClass = {astro-ph.CO},
       adsurl = {https://ui.adsabs.harvard.edu/abs/2017ARA&A..55..343B}
}

@article{Genina_2024,
    author = {Genina, Anna and Springel, Volker and Rantala, Antti},
    title = {A calibrated model for N-body dynamical friction acting on supermassive black holes},
    journal = {Monthly Notices of the Royal Astronomical Society},
    volume = {534},
    number = {1},
    pages = {957-977},
    year = {2024},
    month = {10},
    issn = {0035-8711},
    doi = {10.1093/mnras/stae2144},
    url = {https://doi.org/10.1093/mnras/stae2144},
    eprint = {https://academic.oup.com/mnras/article-pdf/534/1/957/59318132/stae2144.pdf},
}

@article{Kazantzidis_2005,
doi = {10.1086/430139},
url = {https://doi.org/10.1086/430139},
year = {2005},
month = {mar},
publisher = {},
volume = {623},
number = {2},
pages = {L67},
author = {Kazantzidis, Stelios and Mayer, Lucio and Colpi, Monica and Madau, Piero and Debattista, Victor P. and Wadsley, James and Stadel, Joachim and Quinn, Thomas and Moore, Ben},
title = {The Fate of Supermassive Black Holes and the Evolution of the MBH-σ Relation in Merging Galaxies: The Effect of Gaseous Dissipation},
journal = {The Astrophysical Journal}
}

@article{Bondi_1952,
    author = {Bondi, H.},
    title = {On Spherically Symmetrical Accretion},
    journal = {Monthly Notices of the Royal Astronomical Society},
    volume = {112},
    number = {2},
    pages = {195-204},
    year = {1952},
    month = {04},
    issn = {0035-8711},
    doi = {10.1093/mnras/112.2.195},
    url = {https://doi.org/10.1093/mnras/112.2.195},
    eprint = {https://academic.oup.com/mnras/article-pdf/112/2/195/9073555/mnras112-0195.pdf},
}

@ARTICLE{portail+2017,
       author = {{Portail}, Matthieu and {Gerhard}, Ortwin and {Wegg}, Christopher and {Ness}, Melissa},
        title = "{Dynamical modelling of the galactic bulge and bar: the Milky Way's pattern speed, stellar and dark matter mass distribution}",
      journal = {\mnras},
         year = 2017,
        month = feb,
       volume = {465},
       number = {2},
        pages = {1621-1644},
          doi = {10.1093/mnras/stw2819},
archivePrefix = {arXiv},
       eprint = {1608.07954},
 primaryClass = {astro-ph.GA},
       adsurl = {https://ui.adsabs.harvard.edu/abs/2017MNRAS.465.1621P}
}

@ARTICLE{Zhang_2017,
       author = {{Zhang}, XiaoXia and {Lu}, YouJun},
        title = "{On the mean radiative efficiency of accreting massive black holes in AGNs and QSOs}",
      journal = {Science China Physics, Mechanics, and Astronomy},
         year = 2017,
        month = oct,
       volume = {60},
       number = {10},
          eid = {109511},
        pages = {109511},
          doi = {10.1007/s11433-017-9062-1},
archivePrefix = {arXiv},
       eprint = {1902.08332},
 primaryClass = {astro-ph.HE},
       adsurl = {https://ui.adsabs.harvard.edu/abs/2017SCPMA..6009511Z}
}

@ARTICLE{Volonteri_2010,
       author = {{Volonteri}, Marta},
        title = "{Formation of supermassive black holes}",
      journal = {\aapr},
         year = 2010,
        month = jul,
       volume = {18},
       number = {3},
        pages = {279-315},
          doi = {10.1007/s00159-010-0029-x},
archivePrefix = {arXiv},
       eprint = {1003.4404},
 primaryClass = {astro-ph.CO},
       adsurl = {https://ui.adsabs.harvard.edu/abs/2010A&ARv..18..279V}
}

@ARTICLE{Borbolato_2026,
       author = {{Borbolato}, Lais and {Amarante}, Jo{\~a}o A.~S. and {Perottoni}, H{\'e}lio D. and {Rossi}, Silvia and {Debattista}, Victor P. and {Li}, Zhao-Yu and {Deg}, Nathan and {Khachaturyants}, Tigran},
        title = "{The Low-$α$ Splash Population in the Milky Way}",
      journal = {arXiv e-prints},
         year = 2026,
        month = may,
          eid = {arXiv:2605.16050},
        pages = {arXiv:2605.16050},
          doi = {10.48550/arXiv.2605.16050},
archivePrefix = {arXiv},
       eprint = {2605.16050},
 primaryClass = {astro-ph.GA},
       adsurl = {https://ui.adsabs.harvard.edu/abs/2026arXiv260516050B}
}

@misc{Micic_2026,
      title={Feeding The Little Monster: An Accreting Intermediate-Mass Black Hole In A Minor Merger Secondary Galaxy}, 
      author={Marko Mićić and Julia St. Pierre},
      year={2026},
      eprint={2609.02667},
      archivePrefix={arXiv},
      primaryClass={astro-ph.HE},
      url={https://arxiv.org/abs/2609.02667}, 
}

@ARTICLE{Amarante_2025_GASTROII,
       author = {{Amarante}, Jo{\~a}o A.~S. and {Laporte}, Chervin F.~P. and {Debattista}, Victor P. and {Beraldo e Silva}, Leandro and {Limberg}, Guilherme and {Perottoni}, H{\'e}lio D. and {Li}, Zhao-Yu and {Borbolato}, Lais and {Fiteni}, Karl and {Cao}, Chengye and {Deg}, Nathan and {Khachaturyants}, Tigran and {Liao}, Xiaojie},
        title = "{GASTRO Library. II. Exploring Chemical Bimodalities in Disk Galaxies with GSE-like Mergers and Massive Star-forming Clumps}",
      journal = {\apj},
         year = 2026,
        month = jun,
       volume = {1004},
       number = {2},
          eid = {242},
        pages = {242},
          doi = {10.3847/1538-4357/ae70fe},
archivePrefix = {arXiv},
       eprint = {2512.01293},
 primaryClass = {astro-ph.GA},
       adsurl = {https://ui.adsabs.harvard.edu/abs/2026ApJ..1004..242A}
}

@ARTICLE{naidu2021,
       author = {{Naidu}, Rohan P. and {Conroy}, Charlie and {Bonaca}, Ana and {Zaritsky}, Dennis and {Weinberger}, Rainer and {Ting}, Yuan-Sen and {Caldwell}, Nelson and {Tacchella}, Sandro and {Han}, Jiwon Jesse and {Speagle}, Joshua S. and {Cargile}, Phillip A.},
        title = "{Reconstructing the Last Major Merger of the Milky Way with the H3 Survey}",
      journal = {\apj},
         year = 2021,
        month = dec,
       volume = {923},
       number = {1},
          eid = {92},
        pages = {92},
          doi = {10.3847/1538-4357/ac2d2d},
archivePrefix = {arXiv},
       eprint = {2103.03251},
 primaryClass = {astro-ph.GA},
       adsurl = {https://ui.adsabs.harvard.edu/abs/2021ApJ...923...92N}
}

@article{Cho_2026,
doi = {10.3847/2041-8213/ae61ae},
url = {https://doi.org/10.3847/2041-8213/ae61ae},
year = {2026},
month = {may},
publisher = {The American Astronomical Society},
volume = {1002},
number = {2},
pages = {L43},
author = {Cho, Hyerin and Prather, Cora and Narayan, Ramesh and Su, Kung-Yi and Ricarte, Angelo and Natarajan, Priyamvada and Porras-Valverde, Antonio J.},
title = {Bridging Scales in Black Hole Accretion and Feedback: Subgrid Prescription from First Principles},
journal = {The Astrophysical Journal Letters}
}

@ARTICLE{Gnedin_2004,
       author = {{Gnedin}, Oleg Y. and {Kravtsov}, Andrey V. and {Klypin}, Anatoly A. and {Nagai}, Daisuke},
        title = "{Response of Dark Matter Halos to Condensation of Baryons: Cosmological Simulations and Improved Adiabatic Contraction Model}",
      journal = {\apj},
         year = 2004,
        month = nov,
       volume = {616},
       number = {1},
        pages = {16-26},
          doi = {10.1086/424914},
archivePrefix = {arXiv},
       eprint = {astro-ph/0406247},
 primaryClass = {astro-ph},
       adsurl = {https://ui.adsabs.harvard.edu/abs/2004ApJ...616...16G}
}

@ARTICLE{Blumenthal_1986,
       author = {{Blumenthal}, G.~R. and {Faber}, S.~M. and {Flores}, R. and {Primack}, J.~R.},
        title = "{Contraction of Dark Matter Galactic Halos Due to Baryonic Infall}",
      journal = {\apj},
         year = 1986,
        month = feb,
       volume = {301},
        pages = {27},
          doi = {10.1086/163867},
       adsurl = {https://ui.adsabs.harvard.edu/abs/1986ApJ...301...27B}
}

@article{Habouzit_2017,
    author = {Habouzit, Mélanie and Volonteri, Marta and Dubois, Yohan},
    title = {Blossoms from black hole seeds: properties and early growth regulated by supernova feedback},
    journal = {Monthly Notices of the Royal Astronomical Society},
    volume = {468},
    number = {4},
    pages = {3935-3948},
    year = {2017},
    month = {07},
    issn = {0035-8711},
    doi = {10.1093/mnras/stx666},
    url = {https://doi.org/10.1093/mnras/stx666},
    eprint = {https://academic.oup.com/mnras/article-pdf/468/4/3935/13944111/stx666.pdf},
}

@ARTICLE{Dubois_2015,
       author = {{Dubois}, Yohan and {Volonteri}, Marta and {Silk}, Joseph and {Devriendt}, Julien and {Slyz}, Adrianne and {Teyssier}, Romain},
        title = "{Black hole evolution - I. Supernova-regulated black hole growth}",
      journal = {\mnras},
         year = 2015,
        month = sep,
       volume = {452},
       number = {2},
        pages = {1502-1518},
          doi = {10.1093/mnras/stv1416},
archivePrefix = {arXiv},
       eprint = {1504.00018},
 primaryClass = {astro-ph.GA},
       adsurl = {https://ui.adsabs.harvard.edu/abs/2015MNRAS.452.1502D}
}

@article{Pontzen_2017,
    author = {Pontzen, Andrew and Tremmel, Michael and Roth, Nina and Peiris, Hiranya V. and Saintonge, Amélie and Volonteri, Marta and Quinn, Tom and Governato, Fabio},
    title = {How to quench a galaxy},
    journal = {Monthly Notices of the Royal Astronomical Society},
    volume = {465},
    number = {1},
    pages = {547-558},
    year = {2017},
    month = {02},
    issn = {0035-8711},
    doi = {10.1093/mnras/stw2627},
    url = {https://doi.org/10.1093/mnras/stw2627},
    eprint = {https://academic.oup.com/mnras/article-pdf/465/1/547/8592963/stw2627.pdf},
}

@article{Tremmel_2015,
    author = {Tremmel, M. and Governato, F. and Volonteri, M. and Quinn, T. R.},
    title = {Off the beaten path: a new approach to realistically model the orbital decay of supermassive black holes in galaxy formation simulations},
    journal = {Monthly Notices of the Royal Astronomical Society},
    volume = {451},
    number = {2},
    pages = {1868-1874},
    year = {2015},
    month = {06},
    issn = {0035-8711},
    doi = {10.1093/mnras/stv1060},
    url = {https://doi.org/10.1093/mnras/stv1060},
    eprint = {https://academic.oup.com/mnras/article-pdf/451/2/1868/5711066/stv1060.pdf},
}

@article{Shen_2010,
    author = {Shen, S. and Wadsley, J. and Stinson, G.},
    title = {The enrichment of the intergalactic medium with adiabatic feedback – I. Metal cooling and metal diffusion},
    journal = {Monthly Notices of the Royal Astronomical Society},
    volume = {407},
    number = {3},
    pages = {1581-1596},
    year = {2010},
    month = {09},
    issn = {0035-8711},
    doi = {10.1111/j.1365-2966.2010.17047.x},
    url = {https://doi.org/10.1111/j.1365-2966.2010.17047.x},
    eprint = {https://academic.oup.com/mnras/article-pdf/407/3/1581/18434379/mnras0407-1581.pdf},
}

@ARTICLE{Montalban_2021,
       author = {{Montalb{\'a}n}, Josefina and {Mackereth}, J. Ted and {Miglio}, Andrea and {Vincenzo}, Fiorenzo and {Chiappini}, Cristina and {Buldgen}, Gael and {Mosser}, Beno{\^\i}t and {Noels}, Arlette and {Scuflaire}, Richard and {Vrard}, Mathieu and {Willett}, Emma and {Davies}, Guy R. and {Hall}, Oliver J. and {Nielsen}, Martin Bo and {Khan}, Saniya and {Rendle}, Ben M. and {van Rossem}, Walter E. and {Ferguson}, Jason W. and {Chaplin}, William J.},
        title = "{Chronologically dating the early assembly of the Milky Way}",
      journal = {Nature Astronomy},
         year = 2021,
        month = jan,
       volume = {5},
        pages = {640-647},
          doi = {10.1038/s41550-021-01347-7},
archivePrefix = {arXiv},
       eprint = {2006.01783},
 primaryClass = {astro-ph.GA},
       adsurl = {https://ui.adsabs.harvard.edu/abs/2021NatAs...5..640M}
}

@ARTICLE{Massari_2019,
       author = {{Massari}, D. and {Koppelman}, H.~H. and {Helmi}, A.},
        title = "{Origin of the system of globular clusters in the Milky Way}",
      journal = {\aap},
         year = 2019,
        month = oct,
       volume = {630},
          eid = {L4},
        pages = {L4},
          doi = {10.1051/0004-6361/201936135},
archivePrefix = {arXiv},
       eprint = {1906.08271},
 primaryClass = {astro-ph.GA},
       adsurl = {https://ui.adsabs.harvard.edu/abs/2019A&A...630L...4M}
}

@ARTICLE{Callingham_2022,
       author = {{Callingham}, Thomas M. and {Cautun}, Marius and {Deason}, Alis J. and {Frenk}, Carlos S. and {Grand}, Robert J.~J. and {Marinacci}, Federico},
        title = "{The chemo-dynamical groups of Galactic globular clusters}",
      journal = {\mnras},
         year = 2022,
        month = jul,
       volume = {513},
       number = {3},
        pages = {4107-4129},
          doi = {10.1093/mnras/stac1145},
archivePrefix = {arXiv},
       eprint = {2202.00591},
 primaryClass = {astro-ph.GA},
       adsurl = {https://ui.adsabs.harvard.edu/abs/2022MNRAS.513.4107C}
}

@article{Bullock_2001,
doi = {10.1086/321477},
url = {https://dx.doi.org/10.1086/321477},
year = {2001},
month = {jul},
publisher = {},
volume = {555},
number = {1},
pages = {240},
author = {Bullock, J. S. and Dekel, A. and Kolatt, T. S. and Kravtsov, A. V. and Klypin, A. A. and Porciani, C. and Primack, J. R.},
title = {A Universal Angular Momentum Profile for Galactic
Halos},
journal = {The Astrophysical Journal}
}

@ARTICLE{Debattista_2023,
       author = {{Debattista}, Victor P. and {Liddicott}, David J. and {Gonzalez}, Oscar A. and {Beraldo e Silva}, Leandro and {Amarante}, Jo{\~a}o A.~S. and {Lazar}, Ilin and {Zoccali}, Manuela and {Valenti}, Elena and {Fisher}, Deanne B. and {Khachaturyants}, Tigran and {Nidever}, David L. and {Quinn}, Thomas R. and {Du}, Min and {Kassin}, Susan},
        title = "{The Imprint of Clump Formation at High Redshift. II. The Chemistry of the Bulge}",
      journal = {\apj},
         year = 2023,
        month = apr,
       volume = {946},
       number = {2},
          eid = {118},
        pages = {118},
          doi = {10.3847/1538-4357/acbb00},
archivePrefix = {arXiv},
       eprint = {2303.08265},
 primaryClass = {astro-ph.GA},
       adsurl = {https://ui.adsabs.harvard.edu/abs/2023ApJ...946..118D}
}

@article{Belokurov_2018,
    author = {Belokurov, V and Erkal, D and Evans, N W and Koposov, S E and Deason, A J},
    title = {Co-formation of the disc and the stellar halo$*$},
    journal = {Monthly Notices of the Royal Astronomical Society},
    volume = {478},
    number = {1},
    pages = {611-619},
    year = {2018},
    month = {06},
    issn = {0035-8711},
    doi = {10.1093/mnras/sty982},
    url = {https://doi.org/10.1093/mnras/sty982},
    eprint = {https://academic.oup.com/mnras/article-pdf/478/1/611/25005885/sty982.pdf},
}

@article{Orkney_2026,
    author = {Orkney, Matthew D A and Laporte, Chervin F P},
    title = {Build-up and survival of the disc: from numerical models of galaxy formation to the Milky Way},
    journal = {Monthly Notices of the Royal Astronomical Society},
    volume = {548},
    number = {4},
    pages = {staf2154},
    year = {2026},
    month = {06},
    issn = {0035-8711},
    doi = {10.1093/mnras/staf2154},
    url = {https://doi.org/10.1093/mnras/staf2154},
    eprint = {https://academic.oup.com/mnras/article-pdf/548/4/staf2154/68213339/staf2154.pdf},
}

@ARTICLE{EHTC_2022_V,
       author = {{Event Horizon Telescope Collaboration} and {Akiyama}, Kazunori and {Alberdi}, Antxon and {Alef}, Walter and {Algaba}, Juan Carlos and {Anantua}, Richard and {Asada}, Keiichi and {Azulay}, Rebecca and {Bach}, Uwe and {Baczko}, Anne-Kathrin and {Ball}, David and {Balokovi{\'c}}, Mislav and {Barrett}, John and {Baub{\"o}ck}, Michi and {Benson}, Bradford A. and {Bintley}, Dan and {Blackburn}, Lindy and {Blundell}, Raymond and {Bouman}, Katherine L. and {Bower}, Geoffrey C. and {Boyce}, Hope and {Bremer}, Michael and {Brinkerink}, Christiaan D. and {Brissenden}, Roger and {Britzen}, Silke and {Broderick}, Avery E. and {Broguiere}, Dominique and {Bronzwaer}, Thomas and {Bustamante}, Sandra and {Byun}, Do-Young and {Carlstrom}, John E. and {Ceccobello}, Chiara and {Chael}, Andrew and {Chan}, Chi-kwan and {Chatterjee}, Koushik and {Chatterjee}, Shami and {Chen}, Ming-Tang and {Chen}, Yongjun and {Cheng}, Xiaopeng and {Cho}, Ilje and {Christian}, Pierre and {Conroy}, Nicholas S. and {Conway}, John E. and {Cordes}, James M. and {Crawford}, Thomas M. and {Crew}, Geoffrey B. and {Cruz-Osorio}, Alejandro and {Cui}, Yuzhu and {Davelaar}, Jordy and {De Laurentis}, Mariafelicia and {Deane}, Roger and {Dempsey}, Jessica and {Desvignes}, Gregory and {Dexter}, Jason and {Dhruv}, Vedant and {Doeleman}, Sheperd S. and {Dougal}, Sean and {Dzib}, Sergio A. and {Eatough}, Ralph P. and {Emami}, Razieh and {Falcke}, Heino and {Farah}, Joseph and {Fish}, Vincent L. and {Fomalont}, Ed and {Ford}, H. Alyson and {Fraga-Encinas}, Raquel and {Freeman}, William T. and {Friberg}, Per and {Fromm}, Christian M. and {Fuentes}, Antonio and {Galison}, Peter and {Gammie}, Charles F. and {Garc{\'\i}a}, Roberto and {Gentaz}, Olivier and {Georgiev}, Boris and {Goddi}, Ciriaco and {Gold}, Roman and {G{\'o}mez-Ruiz}, Arturo I. and {G{\'o}mez}, Jos{\'e} L. and {Gu}, Minfeng and {Gurwell}, Mark and {Hada}, Kazuhiro and {Haggard}, Daryl and {Haworth}, Kari and {Hecht}, Michael H. and {Hesper}, Ronald and {Heumann}, Dirk and {Ho}, Luis C. and {Ho}, Paul and {Honma}, Mareki and {Huang}, Chih-Wei L. and {Huang}, Lei and {Hughes}, David H. and {Ikeda}, Shiro and {Violette Impellizzeri}, C.~M. and {Inoue}, Makoto and {Issaoun}, Sara and {James}, David J. and {Jannuzi}, Buell T. and {Janssen}, Michael and {Jeter}, Britton and {Jiang}, Wu and {Jim{\'e}nez-Rosales}, Alejandra and {Johnson}, Michael D. and {Jorstad}, Svetlana and {Joshi}, Abhishek V. and {Jung}, Taehyun and {Karami}, Mansour and {Karuppusamy}, Ramesh and {Kawashima}, Tomohisa and {Keating}, Garrett K. and {Kettenis}, Mark and {Kim}, Dong-Jin and {Kim}, Jae-Young and {Kim}, Jongsoo and {Kim}, Junhan and {Kino}, Motoki and {Koay}, Jun Yi and {Kocherlakota}, Prashant and {Kofuji}, Yutaro and {Koch}, Patrick M. and {Koyama}, Shoko and {Kramer}, Carsten and {Kramer}, Michael and {Krichbaum}, Thomas P. and {Kuo}, Cheng-Yu and {La Bella}, Noemi and {Lauer}, Tod R. and {Lee}, Daeyoung and {Lee}, Sang-Sung and {Leung}, Po Kin and {Levis}, Aviad and {Li}, Zhiyuan and {Lico}, Rocco and {Lindahl}, Greg and {Lindqvist}, Michael and {Lisakov}, Mikhail and {Liu}, Jun and {Liu}, Kuo and {Liuzzo}, Elisabetta and {Lo}, Wen-Ping and {Lobanov}, Andrei P. and {Loinard}, Laurent and {Lonsdale}, Colin J. and {Lu}, Ru-Sen and {Mao}, Jirong and {Marchili}, Nicola and {Markoff}, Sera and {Marrone}, Daniel P. and {Marscher}, Alan P. and {Mart{\'\i}-Vidal}, Iv{\'a}n and {Matsushita}, Satoki and {Matthews}, Lynn D. and {Medeiros}, Lia and {Menten}, Karl M. and {Michalik}, Daniel and {Mizuno}, Izumi and {Mizuno}, Yosuke and {Moran}, James M. and {Moriyama}, Kotaro and {Moscibrodzka}, Monika and {M{\"u}ller}, Cornelia and {Mus}, Alejandro and {Musoke}, Gibwa and {Myserlis}, Ioannis and {Nadolski}, Andrew and {Nagai}, Hiroshi and {Nagar}, Neil M. and {Nakamura}, Masanori and {Narayan}, Ramesh and {Narayanan}, Gopal and {Natarajan}, Iniyan and {Nathanail}, Antonios and {Navarro Fuentes}, Santiago and {Neilsen}, Joey and {Neri}, Roberto and {Ni}, Chunchong and {Noutsos}, Aristeidis and {Nowak}, Michael A. and {Oh}, Junghwan and {Okino}, Hiroki and {Olivares}, H{\'e}ctor and {Ortiz-Le{\'o}n}, Gisela N. and {Oyama}, Tomoaki and {{\"O}zel}, Feryal and {Palumbo}, Daniel C.~M. and {Filippos Paraschos}, Georgios and {Park}, Jongho and {Parsons}, Harriet and {Patel}, Nimesh and {Pen}, Ue-Li and {Pesce}, Dominic W. and {Pi{\'e}tu}, Vincent and {Plambeck}, Richard and {PopStefanija}, Aleksandar and {Porth}, Oliver and {P{\"o}tzl}, Felix M. and {Prather}, Ben and {Preciado-L{\'o}pez}, Jorge A. and {Psaltis}, Dimitrios},
        title = "{First Sagittarius A* Event Horizon Telescope Results. V. Testing Astrophysical Models of the Galactic Center Black Hole}",
      journal = {\apjl},
         year = 2022,
        month = may,
       volume = {930},
       number = {2},
          eid = {L16},
        pages = {L16},
          doi = {10.3847/2041-8213/ac6672},
archivePrefix = {arXiv},
       eprint = {2311.09478},
 primaryClass = {astro-ph.HE},
       adsurl = {https://ui.adsabs.harvard.edu/abs/2022ApJ...930L..16E}
}

@ARTICLE{Helmi_2018,
       author = {{Helmi}, Amina and {Babusiaux}, Carine and {Koppelman}, Helmer H. and {Massari}, Davide and {Veljanoski}, Jovan and {Brown}, Anthony G.~A.},
        title = "{The merger that led to the formation of the Milky Way's inner stellar halo and thick disk}",
      journal = {\nat},
         year = 2018,
        month = oct,
       volume = {563},
       number = {7729},
        pages = {85-88},
          doi = {10.1038/s41586-018-0625-x},
archivePrefix = {arXiv},
       eprint = {1806.06038},
 primaryClass = {astro-ph.GA},
       adsurl = {https://ui.adsabs.harvard.edu/abs/2018Natur.563...85H}
}

@ARTICLE{Wadsley_2004,
       author = {{Wadsley}, J.~W. and {Stadel}, J. and {Quinn}, T.},
        title = "{Gasoline: a flexible, parallel implementation of TreeSPH}",
      journal = {\na},
         year = 2004,
        month = feb,
       volume = {9},
       number = {2},
        pages = {137-158},
          doi = {10.1016/j.newast.2003.08.004},
archivePrefix = {arXiv},
       eprint = {astro-ph/0303521},
 primaryClass = {astro-ph},
       adsurl = {https://ui.adsabs.harvard.edu/abs/2004NewA....9..137W}
}

@article{Kale_1993,
author = {Kale, Laxmikant V. and Krishnan, Sanjeev},
title = {CHARM++: a portable concurrent object oriented system based on C++},
year = {1993},
issue_date = {Oct. 1, 1993},
publisher = {Association for Computing Machinery},
address = {New York, NY, USA},
volume = {28},
number = {10},
issn = {0362-1340},
url = {https://doi.org/10.1145/167962.165874},
doi = {10.1145/167962.165874},
journal = {SIGPLAN Not.},
month = oct,
pages = {91–108},
numpages = {18}
}

@ARTICLE{Menon_2015,
       author = {{Menon}, Harshitha and {Wesolowski}, Lukasz and {Zheng}, Gengbin and {Jetley}, Pritish and {Kale}, Laxmikant and {Quinn}, Thomas and {Governato}, Fabio},
        title = "{Adaptive techniques for clustered N-body cosmological simulations}",
      journal = {Computational Astrophysics and Cosmology},
         year = 2015,
        month = mar,
       volume = {2},
          eid = {1},
        pages = {1},
          doi = {10.1186/s40668-015-0007-9},
archivePrefix = {arXiv},
       eprint = {1409.1929},
 primaryClass = {astro-ph.IM},
       adsurl = {https://ui.adsabs.harvard.edu/abs/2015ComAC...2....1M}
}

@inproceedings{Jetley_2010,
author = {Jetley, Pritish and Wesolowski, Lukasz and Gioachin, Filippo and Kal\'{e}, Laxmikant V. and Quinn, Thomas R.},
title = {Scaling Hierarchical N-body Simulations on GPU Clusters},
year = {2010},
isbn = {9781424475599},
publisher = {IEEE Computer Society},
address = {USA},
url = {https://doi.org/10.1109/SC.2010.49},
doi = {10.1109/SC.2010.49},
booktitle = {Proceedings of the 2010 ACM/IEEE International Conference for High Performance Computing, Networking, Storage and Analysis},
pages = {1–11},
numpages = {11},
series = {SC '10}
}

@INPROCEEDINGS{Jetley_2008,

  author={Jetley, Pritish and Gioachin, Filippo and Mendes, Celso and Kale, Laxmikant V. and Quinn, Thomas},

  booktitle={2008 IEEE International Symposium on Parallel and Distributed Processing}, 

  title={Massively parallel cosmological simulations with ChaNGa}, 

  year={2008},

  volume={},

  number={},

  pages={1-12},

  doi={10.1109/IPDPS.2008.4536319}}

@ARTICLE{Limberg_2022,
       author = {{Limberg}, Guilherme and {Souza}, Stefano O. and {P{\'e}rez-Villegas}, Angeles and {Rossi}, Silvia and {Perottoni}, H{\'e}lio D. and {Santucci}, Rafael M.},
        title = "{Reconstructing the Disrupted Dwarf Galaxy Gaia-Sausage/Enceladus Using Its Stars and Globular Clusters}",
      journal = {\apj},
         year = 2022,
        month = aug,
       volume = {935},
       number = {2},
          eid = {109},
        pages = {109},
          doi = {10.3847/1538-4357/ac8159},
archivePrefix = {arXiv},
       eprint = {2206.10505},
 primaryClass = {astro-ph.GA},
       adsurl = {https://ui.adsabs.harvard.edu/abs/2022ApJ...935..109L}
}

@ARTICLE{Limberg_2024,
       author = {{Limberg}, Guilherme},
        title = "{Black Hole Scaling Relations in the Dwarf-galaxy Regime with Gaia-Sausage/Enceladus and {\ensuremath{\omega}}Centauri}",
      journal = {\apjl},
         year = 2024,
        month = dec,
       volume = {977},
       number = {1},
          eid = {L2},
        pages = {L2},
          doi = {10.3847/2041-8213/ad9384},
archivePrefix = {arXiv},
       eprint = {2411.11251},
 primaryClass = {astro-ph.GA},
       adsurl = {https://ui.adsabs.harvard.edu/abs/2024ApJ...977L...2L}
}

@ARTICLE{Bekki_2003,
       author = {{Bekki}, K. and {Freeman}, K.~C.},
        title = "{Formation of {\ensuremath{\omega}} Centauri from an ancient nucleated dwarf galaxy in the young Galactic disc}",
      journal = {\mnras},
         year = 2003,
        month = dec,
       volume = {346},
       number = {2},
        pages = {L11-L15},
          doi = {10.1046/j.1365-2966.2003.07275.x},
archivePrefix = {arXiv},
       eprint = {astro-ph/0310348},
 primaryClass = {astro-ph},
       adsurl = {https://ui.adsabs.harvard.edu/abs/2003MNRAS.346L..11B}
}

@article{Lee_1999,
title = "Multiple stellar populations in the globular cluster $\omega$ Centauri as tracers of a merger event",
author = "Lee, {Y. W.} and Joo, {J. M.} and Sohn, {Y. J.} and Rey, {S. C.} and Lee, {H. C.} and Walker, {A. R.}",
year = "1999",
month = nov,
day = "4",
doi = "10.1038/46985",
language = "English",
volume = "402",
pages = "55--57",
journal = "Nature",
issn = "0028-0836",
publisher = "Nature Research",
number = "6757",

}

@ARTICLE{Haberle_2024,
       author = {{H{\"a}berle}, Maximilian and {Neumayer}, Nadine and {Seth}, Anil and {Bellini}, Andrea and {Libralato}, Mattia and {Baumgardt}, Holger and {Whitaker}, Matthew and {Dumont}, Antoine and {Alfaro-Cuello}, Mayte and {Anderson}, Jay and {Clontz}, Callie and {Kacharov}, Nikolay and {Kamann}, Sebastian and {Feldmeier-Krause}, Anja and {Milone}, Antonino and {Nitschai}, Maria Selina and {Pechetti}, Renuka and {van de Ven}, Glenn},
        title = "{Fast-moving stars around an intermediate-mass black hole in {\ensuremath{\omega}} Centauri}",
      journal = {\nat},
         year = 2024,
        month = jul,
       volume = {631},
       number = {8020},
        pages = {285-288},
          doi = {10.1038/s41586-024-07511-z},
archivePrefix = {arXiv},
       eprint = {2405.06015},
 primaryClass = {astro-ph.GA},
       adsurl = {https://ui.adsabs.harvard.edu/abs/2024Natur.631..285H}
}

@ARTICLE{Greene_2020,
       author = {{Greene}, Jenny E. and {Strader}, Jay and {Ho}, Luis C.},
        title = "{Intermediate-Mass Black Holes}",
      journal = {\araa},
         year = 2020,
        month = aug,
       volume = {58},
        pages = {257-312},
          doi = {10.1146/annurev-astro-032620-021835},
archivePrefix = {arXiv},
       eprint = {1911.09678},
 primaryClass = {astro-ph.GA},
       adsurl = {https://ui.adsabs.harvard.edu/abs/2020ARA&A..58..257G}
}

@ARTICLE{Shakura_1973,
       author = {{Shakura}, N.~I. and {Sunyaev}, R.~A.},
        title = "{Black holes in binary systems. Observational appearance.}",
      journal = {\aap},
         year = 1973,
        month = jan,
       volume = {24},
        pages = {337-355},
       adsurl = {https://ui.adsabs.harvard.edu/abs/1973A&A....24..337S}
}

@ARTICLE{Amarante_2020,
       author = {{Amarante}, Jo{\~a}o A.~S. and {Beraldo e Silva}, Leandro and {Debattista}, Victor P. and {Smith}, Martin C.},
        title = "{The Splash without a Merger}",
      journal = {\apjl},
         year = 2020,
        month = mar,
       volume = {891},
       number = {2},
          eid = {L30},
        pages = {L30},
          doi = {10.3847/2041-8213/ab78a4},
archivePrefix = {arXiv},
       eprint = {1912.12690},
 primaryClass = {astro-ph.GA},
       adsurl = {https://ui.adsabs.harvard.edu/abs/2020ApJ...891L..30A}
}

@ARTICLE{Navarro_1997,
       author = {{Navarro}, Julio F. and {Frenk}, Carlos S. and {White}, Simon D.~M.},
        title = "{A Universal Density Profile from Hierarchical Clustering}",
      journal = {\apj},
         year = 1997,
        month = dec,
       volume = {490},
       number = {2},
        pages = {493-508},
          doi = {10.1086/304888},
archivePrefix = {arXiv},
       eprint = {astro-ph/9611107},
 primaryClass = {astro-ph},
       adsurl = {https://ui.adsabs.harvard.edu/abs/1997ApJ...490..493N}
}

@ARTICLE{Beraldo_Silva_2021,
       author = {{Beraldo e Silva}, Leandro and {Debattista}, Victor P. and {Nidever}, David and {Amarante}, Jo{\~a}o A.~S. and {Garver}, Bethany},
        title = "{Co-formation of the thin and thick discs revealed by APOGEE-DR16 and Gaia-DR2}",
      journal = {\mnras},
         year = 2021,
        month = mar,
       volume = {502},
       number = {1},
        pages = {260-272},
          doi = {10.1093/mnras/staa3966},
archivePrefix = {arXiv},
       eprint = {2009.03346},
 primaryClass = {astro-ph.GA},
       adsurl = {https://ui.adsabs.harvard.edu/abs/2021MNRAS.502..260B}
}

@ARTICLE{Widrow_2008,
       author = {{Widrow}, Lawrence M. and {Pym}, Brent and {Dubinski}, John},
        title = "{Dynamical Blueprints for Galaxies}",
      journal = {\apj},
         year = 2008,
        month = jun,
       volume = {679},
       number = {2},
        pages = {1239-1259},
          doi = {10.1086/587636},
archivePrefix = {arXiv},
       eprint = {0801.3414},
 primaryClass = {astro-ph},
       adsurl = {https://ui.adsabs.harvard.edu/abs/2008ApJ...679.1239W}
}

@ARTICLE{Widrow_2005,
       author = {{Widrow}, Lawrence M. and {Dubinski}, John},
        title = "{Equilibrium Disk-Bulge-Halo Models for the Milky Way and Andromeda Galaxies}",
      journal = {\apj},
         year = 2005,
        month = oct,
       volume = {631},
       number = {2},
        pages = {838-855},
          doi = {10.1086/432710},
archivePrefix = {arXiv},
       eprint = {astro-ph/0506177},
 primaryClass = {astro-ph},
       adsurl = {https://ui.adsabs.harvard.edu/abs/2005ApJ...631..838W}
}

@ARTICLE{Kuijken_1995,
       author = {{Kuijken}, K. and {Dubinski}, J.},
        title = "{Nearly Self-Consistent Disc / Bulge / Halo Models for Galaxies}",
      journal = {\mnras},
         year = 1995,
        month = dec,
       volume = {277},
        pages = {1341},
          doi = {10.1093/mnras/277.4.1341},
       adsurl = {https://ui.adsabs.harvard.edu/abs/1995MNRAS.277.1341K}
}

@misc{Petersson_2025,
      title={The Noctua Suite of Simulations -- The Difficulty of Growing Massive Black Holes in Low-Mass Dwarf Galaxies}, 
      author={Jonathan Petersson and Michaela Hirschmann and Robin G. Tress and Marion Farcy and Simon C. O. Glover and Ralf S. Klessen and Thorsten Naab and Christian Partmann and David J. Whitworth},
      year={2025},
      eprint={2504.08035},
      archivePrefix={arXiv},
      primaryClass={astro-ph.GA},
      url={https://arxiv.org/abs/2504.08035}, 
}

@article{Booth_Schaye_2009,
    author = {Booth, C. M. and Schaye, Joop},
    title = {Cosmological simulations of the growth of supermassive black holes and feedback from active galactic nuclei: method and tests},
    journal = {Monthly Notices of the Royal Astronomical Society},
    volume = {398},
    number = {1},
    pages = {53-74},
    year = {2009},
    month = {08},
    issn = {0035-8711},
    doi = {10.1111/j.1365-2966.2009.15043.x},
    url = {https://doi.org/10.1111/j.1365-2966.2009.15043.x},
    eprint = {https://academic.oup.com/mnras/article-pdf/398/1/53/3815919/mnras0398-0053.pdf},
}

@article{Tremmel_2017,
    author = {Tremmel, M. and Karcher, M. and Governato, F. and Volonteri, M. and Quinn, T. R. and Pontzen, A. and Anderson, L. and Bellovary, J.},
    title = {The Romulus cosmological simulations: a physical approach to the formation, dynamics and accretion models of SMBHs},
    journal = {Monthly Notices of the Royal Astronomical Society},
    volume = {470},
    number = {1},
    pages = {1121-1139},
    year = {2017},
    month = {05},
    issn = {0035-8711},
    doi = {10.1093/mnras/stx1160},
    url = {https://doi.org/10.1093/mnras/stx1160},
    eprint = {https://academic.oup.com/mnras/article-pdf/470/1/1121/17933819/stx1160.pdf},
}

@ARTICLE{Beraldo_Silva_2020,
       author = {{Beraldo e Silva}, Leandro and {Debattista}, Victor P. and {Khachaturyants}, Tigran and {Nidever}, David},
        title = "{Geometric properties of galactic discs with clumpy episodes}",
      journal = {\mnras},
         year = 2020,
        month = mar,
       volume = {492},
       number = {4},
        pages = {4716-4726},
          doi = {10.1093/mnras/staa065},
archivePrefix = {arXiv},
       eprint = {1911.03717},
 primaryClass = {astro-ph.GA},
       adsurl = {https://ui.adsabs.harvard.edu/abs/2020MNRAS.492.4716B}
}

@article{Clarke_2019,
    author = {Clarke, Adam J and Debattista, Victor P and Nidever, David L and Loebman, Sarah R and Simons, Raymond C and Kassin, Susan and Du, Min and Ness, Melissa and Fisher, Deanne B and Quinn, Thomas R and Wadsley, James and Freeman, Ken C and Popescu, Cristina C},
    title = {The imprint of clump formation at high redshift – I. A disc $\alpha$-abundance dichotomy},
    journal = {Monthly Notices of the Royal Astronomical Society},
    volume = {484},
    number = {3},
    pages = {3476-3490},
    year = {2019},
    month = {01},
    issn = {0035-8711},
    doi = {10.1093/mnras/stz104},
    url = {https://doi.org/10.1093/mnras/stz104},
    eprint = {https://academic.oup.com/mnras/article-pdf/484/3/3476/27711724/stz104.pdf},
}

@ARTICLE{Stinson_2006,
       author = {{Stinson}, Greg and {Seth}, Anil and {Katz}, Neal and {Wadsley}, James and {Governato}, Fabio and {Quinn}, Tom},
        title = "{Star formation and feedback in smoothed particle hydrodynamic simulations - I. Isolated galaxies}",
      journal = {\mnras},
         year = 2006,
        month = dec,
       volume = {373},
       number = {3},
        pages = {1074-1090},
          doi = {10.1111/j.1365-2966.2006.11097.x},
archivePrefix = {arXiv},
       eprint = {astro-ph/0602350},
 primaryClass = {astro-ph},
       adsurl = {https://ui.adsabs.harvard.edu/abs/2006MNRAS.373.1074S}
}

@article{Amarante_2022,
doi = {10.3847/1538-4357/ac8b0d},
url = {https://dx.doi.org/10.3847/1538-4357/ac8b0d},
year = {2022},
month = {sep},
publisher = {The American Astronomical Society},
volume = {937},
number = {1},
pages = {12},
author = {Amarante, João A. S. and Debattista, Victor P. and Beraldo E Silva, Leandro and Laporte, Chervin F. P. and Deg, Nathan},
title = {Gastro Library. I. The Simulated Chemodynamical Properties of Several Gaia–Sausage–Enceladus-like Stellar Halos},
journal = {The Astrophysical Journal}
}

@ARTICLE{Kormendy2013blackHoles,
       author = {{Kormendy}, John and {Ho}, Luis C.},
        title = "{Coevolution (Or Not) of Supermassive Black Holes and Host Galaxies}",
      journal = {\araa},
         year = 2013,
        month = aug,
       volume = {51},
       number = {1},
        pages = {511-653},
          doi = {10.1146/annurev-astro-082708-101811},
archivePrefix = {arXiv},
       eprint = {1304.7762},
 primaryClass = {astro-ph.CO},
       adsurl = {https://ui.adsabs.harvard.edu/abs/2013ARA&A..51..511K}
}

@ARTICLE{Inayoshi2020smbh,
       author = {{Inayoshi}, Kohei and {Visbal}, Eli and {Haiman}, Zolt{\'a}n},
        title = "{The Assembly of the First Massive Black Holes}",
      journal = {\araa},
         year = 2020,
        month = aug,
       volume = {58},
        pages = {27-97},
          doi = {10.1146/annurev-astro-120419-014455},
archivePrefix = {arXiv},
       eprint = {1911.05791},
 primaryClass = {astro-ph.GA},
       adsurl = {https://ui.adsabs.harvard.edu/abs/2020ARA&A..58...27I}
}

@ARTICLE{Ghez1998sgrA*,
       author = {{Ghez}, A.~M. and {Klein}, B.~L. and {Morris}, M. and {Becklin}, E.~E.},
        title = "{High Proper-Motion Stars in the Vicinity of Sagittarius A*: Evidence for a Supermassive Black Hole at the Center of Our Galaxy}",
      journal = {\apj},
         year = 1998,
        month = dec,
       volume = {509},
       number = {2},
        pages = {678-686},
          doi = {10.1086/306528},
archivePrefix = {arXiv},
       eprint = {astro-ph/9807210},
 primaryClass = {astro-ph},
       adsurl = {https://ui.adsabs.harvard.edu/abs/1998ApJ...509..678G}
}

@ARTICLE{EckartGenzel1997sgrA*,
       author = {{Eckart}, A. and {Genzel}, R.},
        title = "{Stellar proper motions in the central 0.1 PC of the Galaxy}",
      journal = {\mnras},
         year = 1997,
        month = jan,
       volume = {284},
       number = {3},
        pages = {576-598},
          doi = {10.1093/mnras/284.3.576},
       adsurl = {https://ui.adsabs.harvard.edu/abs/1997MNRAS.284..576E}
}

@ARTICLE{DiMatteo2005AGNfeedback,
       author = {{Di Matteo}, Tiziana and {Springel}, Volker and {Hernquist}, Lars},
        title = "{Energy input from quasars regulates the growth and activity of black holes and their host galaxies}",
      journal = {\nat},
         year = 2005,
        month = feb,
       volume = {433},
       number = {7026},
        pages = {604-607},
          doi = {10.1038/nature03335},
archivePrefix = {arXiv},
       eprint = {astro-ph/0502199},
 primaryClass = {astro-ph},
       adsurl = {https://ui.adsabs.harvard.edu/abs/2005Natur.433..604D}
}

@ARTICLE{silk2017_AGNinDwarfs,
       author = {{Silk}, Joseph},
        title = "{Feedback by Massive Black Holes in Gas-rich Dwarf Galaxies}",
      journal = {\apjl},
         year = 2017,
        month = apr,
       volume = {839},
       number = {1},
          eid = {L13},
        pages = {L13},
          doi = {10.3847/2041-8213/aa67da},
archivePrefix = {arXiv},
       eprint = {1703.08553},
 primaryClass = {astro-ph.GA},
       adsurl = {https://ui.adsabs.harvard.edu/abs/2017ApJ...839L..13S}
}

@ARTICLE{Larson1974_SNfeedbackGals,
       author = {{Larson}, Richard B.},
        title = "{Effects of supernovae on the early evolution of galaxies}",
      journal = {\mnras},
         year = 1974,
        month = nov,
       volume = {169},
        pages = {229-246},
          doi = {10.1093/mnras/169.2.229},
       adsurl = {https://ui.adsabs.harvard.edu/abs/1974MNRAS.169..229L}
}

@ARTICLE{sdssYork,
       author = {{York}, Donald G. and {Adelman}, J. and {Anderson}, John E., Jr. and
         {Anderson}, Scott F. and {Annis}, James and {Bahcall}, Neta A. and
         {Bakken}, J.~A. and {Barkhouser}, Robert and {Bastian}, Steven and
         {Berman}, Eileen and {Boroski}, William N. and {Bracker}, Steve and
         {Briegel}, Charlie and {Briggs}, John W. and {Brinkmann}, J. and
         {Brunner}, Robert and {Burles}, Scott and {Carey}, Larry and
         {Carr}, Michael A. and {Castander}, Francisco J. and {Chen}, Bing and
         {Colestock}, Patrick L. and {Connolly}, A.~J. and {Crocker}, J.~H. and
         {Csabai}, Istv{\'a}n and {Czarapata}, Paul C. and {Davis}, John Eric and
         {Doi}, Mamoru and {Dombeck}, Tom and {Eisenstein}, Daniel and
         {Ellman}, Nancy and {Elms}, Brian R. and {Evans}, Michael L. and
         {Fan}, Xiaohui and {Federwitz}, Glenn R. and {Fiscelli}, Larry and
         {Friedman}, Scott and {Frieman}, Joshua A. and {Fukugita}, Masataka and
         {Gillespie}, Bruce and {Gunn}, James E. and {Gurbani}, Vijay K. and
         {de Haas}, Ernst and {Haldeman}, Merle and {Harris}, Frederick H. and
         {Hayes}, J. and {Heckman}, Timothy M. and {Hennessy}, G.~S. and
         {Hindsley}, Robert B. and {Holm}, Scott and {Holmgren}, Donald J. and
         {Huang}, Chi-hao and {Hull}, Charles and {Husby}, Don and
         {Ichikawa}, Shin-Ichi and {Ichikawa}, Takashi and
         {Ivezi{\'c}}, {\v{Z}}eljko and {Kent}, Stephen and {Kim}, Rita S.~J. and
         {Kinney}, E. and {Klaene}, Mark and {Kleinman}, A.~N. and
         {Kleinman}, S. and {Knapp}, G.~R. and {Korienek}, John and
         {Kron}, Richard G. and {Kunszt}, Peter Z. and {Lamb}, D.~Q. and
         {Lee}, B. and {Leger}, R. French and {Limmongkol}, Siriluk and
         {Lindenmeyer}, Carl and {Long}, Daniel C. and {Loomis}, Craig and
         {Loveday}, Jon and {Lucinio}, Rich and {Lupton}, Robert H. and
         {MacKinnon}, Bryan and {Mannery}, Edward J. and {Mantsch}, P.~M. and
         {Margon}, Bruce and {McGehee}, Peregrine and {McKay}, Timothy A. and
         {Meiksin}, Avery and {Merelli}, Aronne and {Monet}, David G. and
         {Munn}, Jeffrey A. and {Narayanan}, Vijay K. and {Nash}, Thomas and
         {Neilsen}, Eric and {Neswold}, Rich and {Newberg}, Heidi Jo and
         {Nichol}, R.~C. and {Nicinski}, Tom and {Nonino}, Mario and
         {Okada}, Norio and {Okamura}, Sadanori and {Ostriker}, Jeremiah P. and
         {Owen}, Russell and {Pauls}, A. George and {Peoples}, John and
         {Peterson}, R.~L. and {Petravick}, Donald and {Pier}, Jeffrey R. and
         {Pope}, Adrian and {Pordes}, Ruth and {Prosapio}, Angela and
         {Rechenmacher}, Ron and {Quinn}, Thomas R. and {Richards}, Gordon T. and
         {Richmond}, Michael W. and {Rivetta}, Claudio H. and
         {Rockosi}, Constance M. and {Ruthmansdorfer}, Kurt and {Sand
        ford}, Dale and {Schlegel}, David J. and {Schneider}, Donald P. and
         {Sekiguchi}, Maki and {Sergey}, Gary and {Shimasaku}, Kazuhiro and
         {Siegmund}, Walter A. and {Smee}, Stephen and {Smith}, J. Allyn and
         {Snedden}, S. and {Stone}, R. and {Stoughton}, Chris and
         {Strauss}, Michael A. and {Stubbs}, Christopher and {SubbaRao}, Mark and
         {Szalay}, Alexander S. and {Szapudi}, Istvan and {Szokoly}, Gyula P. and
         {Thakar}, Anirudda R. and {Tremonti}, Christy and {Tucker}, Douglas L. and
         {Uomoto}, Alan and {Vanden Berk}, Dan and {Vogeley}, Michael S. and
         {Waddell}, Patrick and {Wang}, Shu-i. and {Watanabe}, Masaru and
         {Weinberg}, David H. and {Yanny}, Brian and {Yasuda}, Naoki and
         {SDSS Collaboration}},
        title = "{The Sloan Digital Sky Survey: Technical Summary}",
      journal = {\aj},
         year = 2000,
        month = sep,
       volume = {120},
       number = {3},
        pages = {1579-1587},
          doi = {10.1086/301513},
archivePrefix = {arXiv},
       eprint = {astro-ph/0006396},
 primaryClass = {astro-ph},
       adsurl = {https://ui.adsabs.harvard.edu/abs/2000AJ....120.1579Y}
}

@ARTICLE{Reines2013bhs,
       author = {{Reines}, Amy E. and {Greene}, Jenny E. and {Geha}, Marla},
        title = "{Dwarf Galaxies with Optical Signatures of Active Massive Black Holes}",
      journal = {\apj},
         year = 2013,
        month = oct,
       volume = {775},
       number = {2},
          eid = {116},
        pages = {116},
          doi = {10.1088/0004-637X/775/2/116},
archivePrefix = {arXiv},
       eprint = {1308.0328},
 primaryClass = {astro-ph.CO},
       adsurl = {https://ui.adsabs.harvard.edu/abs/2013ApJ...775..116R}
}

@ARTICLE{Pucha2025_dwarfAGNsDESI,
       author = {{Pucha}, Ragadeepika and {Juneau}, S. and {Dey}, Arjun and {Siudek}, M. and {Mezcua}, M. and {Moustakas}, J. and {BenZvi}, S. and {Hainline}, K. and {Hviding}, R. and {Mao}, Yao-Yuan and {Alexander}, D.~M. and {Alfarsy}, R. and {Circosta}, C. and {Guo}, Wei-Jian and {Manwadkar}, V. and {Martini}, P. and {Weaver}, B.~A. and {Aguilar}, J. and {Ahlen}, S. and {Bianchi}, D. and {Brooks}, D. and {Canning}, R. and {Claybaugh}, T. and {Dawson}, K. and {de la Macorra}, A. and {Dey}, Biprateep and {Doel}, P. and {Font-Ribera}, A. and {Forero-Romero}, J.~E. and {Gazta{\~n}aga}, E. and {Gontcho A Gontcho}, S. and {Gutierrez}, G. and {Honscheid}, K. and {Kehoe}, R. and {Koposov}, S.~E. and {Lambert}, A. and {Landriau}, M. and {Le Guillou}, L. and {Meisner}, A. and {Miquel}, R. and {Prada}, F. and {Rossi}, G. and {Sanchez}, E. and {Schlegel}, D. and {Schubnell}, M. and {Seo}, H. and {Sprayberry}, D. and {Tarl{\'e}}, G. and {Zou}, H.},
        title = "{Tripling the Census of Dwarf AGN Candidates Using DESI Early Data}",
      journal = {\apj},
         year = 2025,
        month = mar,
       volume = {982},
       number = {1},
          eid = {10},
        pages = {10},
          doi = {10.3847/1538-4357/adb1dd},
archivePrefix = {arXiv},
       eprint = {2411.00091},
 primaryClass = {astro-ph.GA},
       adsurl = {https://ui.adsabs.harvard.edu/abs/2025ApJ...982...10P}
}

@ARTICLE{BPTdiagram1981,
       author = {{Baldwin}, J.~A. and {Phillips}, M.~M. and {Terlevich}, R.},
        title = "{Classification parameters for the emission-line spectra of extragalactic objects.}",
      journal = {\pasp},
         year = 1981,
        month = feb,
       volume = {93},
        pages = {5-19},
          doi = {10.1086/130766},
       adsurl = {https://ui.adsabs.harvard.edu/abs/1981PASP...93....5B}
}

@ARTICLE{Mannucci2010mzr,
       author = {{Mannucci}, F. and {Cresci}, G. and {Maiolino}, R. and {Marconi}, A. and {Gnerucci}, A.},
        title = "{A fundamental relation between mass, star formation rate and metallicity in local and high-redshift galaxies}",
      journal = {\mnras},
         year = 2010,
        month = nov,
       volume = {408},
       number = {4},
        pages = {2115-2127},
          doi = {10.1111/j.1365-2966.2010.17291.x},
archivePrefix = {arXiv},
       eprint = {1005.0006},
 primaryClass = {astro-ph.CO},
       adsurl = {https://ui.adsabs.harvard.edu/abs/2010MNRAS.408.2115M}
}

@ARTICLE{Tremonti2004mzr,
       author = {{Tremonti}, Christy A. and {Heckman}, Timothy M. and {Kauffmann}, Guinevere and {Brinchmann}, Jarle and {Charlot}, St{\'e}phane and {White}, Simon D.~M. and {Seibert}, Mark and {Peng}, Eric W. and {Schlegel}, David J. and {Uomoto}, Alan and {Fukugita}, Masataka and {Brinkmann}, Jon},
        title = "{The Origin of the Mass-Metallicity Relation: Insights from 53,000 Star-forming Galaxies in the Sloan Digital Sky Survey}",
      journal = {\apj},
         year = 2004,
        month = oct,
       volume = {613},
       number = {2},
        pages = {898-913},
          doi = {10.1086/423264},
archivePrefix = {arXiv},
       eprint = {astro-ph/0405537},
 primaryClass = {astro-ph},
       adsurl = {https://ui.adsabs.harvard.edu/abs/2004ApJ...613..898T}
}

@ARTICLE{kirby2013mzr,
       author = {{Kirby}, Evan N. and {Cohen}, Judith G. and {Guhathakurta}, Puragra and {Cheng}, Lucy and {Bullock}, James S. and {Gallazzi}, Anna},
        title = "{The Universal Stellar Mass-Stellar Metallicity Relation for Dwarf Galaxies}",
      journal = {\apj},
         year = 2013,
        month = dec,
       volume = {779},
       number = {2},
          eid = {102},
        pages = {102},
          doi = {10.1088/0004-637X/779/2/102},
archivePrefix = {arXiv},
       eprint = {1310.0814},
 primaryClass = {astro-ph.GA},
       adsurl = {https://ui.adsabs.harvard.edu/abs/2013ApJ...779..102K}
}

@ARTICLE{Mezcua2017review,
       author = {{Mezcua}, Mar},
        title = "{Observational evidence for intermediate-mass black holes}",
      journal = {International Journal of Modern Physics D},
         year = 2017,
        month = jan,
       volume = {26},
       number = {11},
          eid = {1730021},
        pages = {1730021},
          doi = {10.1142/S021827181730021X},
archivePrefix = {arXiv},
       eprint = {1705.09667},
 primaryClass = {astro-ph.GA},
       adsurl = {https://ui.adsabs.harvard.edu/abs/2017IJMPD..2630021M}
}

@ARTICLE{Groves2006metalpoorAGN,
       author = {{Groves}, Brent A. and {Heckman}, Timothy M. and {Kauffmann}, Guinevere},
        title = "{Emission-line diagnostics of low-metallicity active galactic nuclei}",
      journal = {\mnras},
         year = 2006,
        month = oct,
       volume = {371},
       number = {4},
        pages = {1559-1569},
          doi = {10.1111/j.1365-2966.2006.10812.x},
archivePrefix = {arXiv},
       eprint = {astro-ph/0607311},
 primaryClass = {astro-ph},
       adsurl = {https://ui.adsabs.harvard.edu/abs/2006MNRAS.371.1559G}
}

@ARTICLE{WasleskeBaldassare2024dwarfAGNs,
       author = {{Wasleske}, Erik J. and {Baldassare}, Vivienne F.},
        title = "{Active Dwarf Galaxy Database. I. Overlap between Active Galactic Nuclei Selected by Different Techniques}",
      journal = {\apj},
         year = 2024,
        month = aug,
       volume = {971},
       number = {1},
          eid = {68},
        pages = {68},
          doi = {10.3847/1538-4357/ad5442},
archivePrefix = {arXiv},
       eprint = {2405.20312},
 primaryClass = {astro-ph.GA},
       adsurl = {https://ui.adsabs.harvard.edu/abs/2024ApJ...971...68W}
}

@ARTICLE{burke2025dwarfAGN,
       author = {{Burke}, Colin J. and {Natarajan}, Priyamvada and {Baldassare}, Vivienne F. and {Geha}, Marla},
        title = "{Multiwavelength Constraints on the Local Black Hole Occupation Fraction}",
      journal = {\apj},
         year = 2025,
        month = jan,
       volume = {978},
       number = {1},
          eid = {77},
        pages = {77},
          doi = {10.3847/1538-4357/ad94d9},
archivePrefix = {arXiv},
       eprint = {2410.11177},
 primaryClass = {astro-ph.GA},
       adsurl = {https://ui.adsabs.harvard.edu/abs/2025ApJ...978...77B}
}

@ARTICLE{zou2025dwarfAGN,
       author = {{Zou}, Fan and {Gallo}, Elena and {Seth}, Anil C. and {Hodges-Kluck}, Edmund and {Ohlson}, David and {Treu}, Tommaso and {Baldassare}, Vivienne F. and {Brandt}, W.~N. and {Greene}, Jenny E. and {Madau}, Piero and {Nguyen}, Dieu D. and {Plotkin}, Richard M. and {Reines}, Amy E. and {Sesana}, Alberto and {Woo}, Jong-Hak and {Wu}, Jianfeng},
        title = "{Central Massive Black Holes Are Not Ubiquitous in Local Low-mass Galaxies}",
      journal = {\apj},
         year = 2025,
        month = oct,
       volume = {992},
       number = {2},
          eid = {176},
        pages = {176},
          doi = {10.3847/1538-4357/ae06a1},
archivePrefix = {arXiv},
       eprint = {2510.05252},
 primaryClass = {astro-ph.GA},
       adsurl = {https://ui.adsabs.harvard.edu/abs/2025ApJ...992..176Z}
}

@ARTICLE{Helmi2020,
       author = {{Helmi}, Amina},
        title = "{Streams, Substructures, and the Early History of the Milky Way}",
      journal = {\araa},
         year = 2020,
        month = aug,
       volume = {58},
        pages = {205-256},
          doi = {10.1146/annurev-astro-032620-021917},
archivePrefix = {arXiv},
       eprint = {2002.04340},
 primaryClass = {astro-ph.GA},
       adsurl = {https://ui.adsabs.harvard.edu/abs/2020ARA&A..58..205H}
}

@ARTICLE{WhiteRees1978,
       author = {{White}, S.~D.~M. and {Rees}, M.~J.},
        title = "{Core condensation in heavy halos: a two-stage theory for galaxy formation and clustering.}",
      journal = {\mnras},
         year = 1978,
        month = may,
       volume = {183},
        pages = {341-358},
          doi = {10.1093/mnras/183.3.341},
       adsurl = {https://ui.adsabs.harvard.edu/abs/1978MNRAS.183..341W}
}

@ARTICLE{White1991cdm,
       author = {{White}, Simon D.~M. and {Frenk}, Carlos S.},
        title = "{Galaxy Formation through Hierarchical Clustering}",
      journal = {\apj},
         year = 1991,
        month = sep,
       volume = {379},
        pages = {52},
          doi = {10.1086/170483},
       adsurl = {https://ui.adsabs.harvard.edu/abs/1991ApJ...379...52W}
}

@ARTICLE{BullockJohnston2005,
       author = {{Bullock}, James S. and {Johnston}, Kathryn V.},
        title = "{Tracing Galaxy Formation with Stellar Halos. I. Methods}",
      journal = {\apj},
         year = 2005,
        month = dec,
       volume = {635},
       number = {2},
        pages = {931-949},
          doi = {10.1086/497422},
archivePrefix = {arXiv},
       eprint = {astro-ph/0506467},
 primaryClass = {astro-ph},
       adsurl = {https://ui.adsabs.harvard.edu/abs/2005ApJ...635..931B}
}

@ARTICLE{Johnston1998,
       author = {{Johnston}, Kathryn V.},
        title = "{A Prescription for Building the Milky Way's Halo from Disrupted Satellites}",
      journal = {\apj},
         year = 1998,
        month = mar,
       volume = {495},
       number = {1},
        pages = {297-308},
          doi = {10.1086/305273},
archivePrefix = {arXiv},
       eprint = {astro-ph/9710007},
 primaryClass = {astro-ph},
       adsurl = {https://ui.adsabs.harvard.edu/abs/1998ApJ...495..297J}
}

@ARTICLE{GaiaDR3_2023,
       author = {{Gaia Collaboration} and {Vallenari}, A. and {Brown}, A.~G.~A. and {Prusti}, T. and {de Bruijne}, J.~H.~J. and {Arenou}, F. and {Babusiaux}, C. and {Biermann}, M. and {Creevey}, O.~L. and {Ducourant}, C. and {Evans}, D.~W. and {Eyer}, L. and {Guerra}, R. and {Hutton}, A. and {Jordi}, C. and {Klioner}, S.~A. and {Lammers}, U.~L. and {Lindegren}, L. and {Luri}, X. and {Mignard}, F. and {Panem}, C. and {Pourbaix}, D. and {Randich}, S. and {Sartoretti}, P. and {Soubiran}, C. and {Tanga}, P. and {Walton}, N.~A. and {Bailer-Jones}, C.~A.~L. and {Bastian}, U. and {Drimmel}, R. and {Jansen}, F. and {Katz}, D. and {Lattanzi}, M.~G. and {van Leeuwen}, F. and {Bakker}, J. and {Cacciari}, C. and {Casta{\~n}eda}, J. and {De Angeli}, F. and {Fabricius}, C. and {Fouesneau}, M. and {Fr{\'e}mat}, Y. and {Galluccio}, L. and {Guerrier}, A. and {Heiter}, U. and {Masana}, E. and {Messineo}, R. and {Mowlavi}, N. and {Nicolas}, C. and {Nienartowicz}, K. and {Pailler}, F. and {Panuzzo}, P. and {Riclet}, F. and {Roux}, W. and {Seabroke}, G.~M. and {Sordo}, R. and {Th{\'e}venin}, F. and {Gracia-Abril}, G. and {Portell}, J. and {Teyssier}, D. and {Altmann}, M. and {Andrae}, R. and {Audard}, M. and {Bellas-Velidis}, I. and {Benson}, K. and {Berthier}, J. and {Blomme}, R. and {Burgess}, P.~W. and {Busonero}, D. and {Busso}, G. and {C{\'a}novas}, H. and {Carry}, B. and {Cellino}, A. and {Cheek}, N. and {Clementini}, G. and {Damerdji}, Y. and {Davidson}, M. and {de Teodoro}, P. and {Nu{\~n}ez Campos}, M. and {Delchambre}, L. and {Dell'Oro}, A. and {Esquej}, P. and {Fern{\'a}ndez-Hern{\'a}ndez}, J. and {Fraile}, E. and {Garabato}, D. and {Garc{\'\i}a-Lario}, P. and {Gosset}, E. and {Haigron}, R. and {Halbwachs}, J.-L. and {Hambly}, N.~C. and {Harrison}, D.~L. and {Hern{\'a}ndez}, J. and {Hestroffer}, D. and {Hodgkin}, S.~T. and {Holl}, B. and {Jan{\ss}en}, K. and {Jevardat de Fombelle}, G. and {Jordan}, S. and {Krone-Martins}, A. and {Lanzafame}, A.~C. and {L{\"o}ffler}, W. and {Marchal}, O. and {Marrese}, P.~M. and {Moitinho}, A. and {Muinonen}, K. and {Osborne}, P. and {Pancino}, E. and {Pauwels}, T. and {Recio-Blanco}, A. and {Reyl{\'e}}, C. and {Riello}, M. and {Rimoldini}, L. and {Roegiers}, T. and {Rybizki}, J. and {Sarro}, L.~M. and {Siopis}, C. and {Smith}, M. and {Sozzetti}, A. and {Utrilla}, E. and {van Leeuwen}, M. and {Abbas}, U. and {{\'A}brah{\'a}m}, P. and {Abreu Aramburu}, A. and {Aerts}, C. and {Aguado}, J.~J. and {Ajaj}, M. and {Aldea-Montero}, F. and {Altavilla}, G. and {{\'A}lvarez}, M.~A. and {Alves}, J. and {Anders}, F. and {Anderson}, R.~I. and {Anglada Varela}, E. and {Antoja}, T. and {Baines}, D. and {Baker}, S.~G. and {Balaguer-N{\'u}{\~n}ez}, L. and {Balbinot}, E. and {Balog}, Z. and {Barache}, C. and {Barbato}, D. and {Barros}, M. and {Barstow}, M.~A. and {Bartolom{\'e}}, S. and {Bassilana}, J.-L. and {Bauchet}, N. and {Becciani}, U. and {Bellazzini}, M. and {Berihuete}, A. and {Bernet}, M. and {Bertone}, S. and {Bianchi}, L. and {Binnenfeld}, A. and {Blanco-Cuaresma}, S. and {Blazere}, A. and {Boch}, T. and {Bombrun}, A. and {Bossini}, D. and {Bouquillon}, S. and {Bragaglia}, A. and {Bramante}, L. and {Breedt}, E. and {Bressan}, A. and {Brouillet}, N. and {Brugaletta}, E. and {Bucciarelli}, B. and {Burlacu}, A. and {Butkevich}, A.~G. and {Buzzi}, R. and {Caffau}, E. and {Cancelliere}, R. and {Cantat-Gaudin}, T. and {Carballo}, R. and {Carlucci}, T. and {Carnerero}, M.~I. and {Carrasco}, J.~M. and {Casamiquela}, L. and {Castellani}, M. and {Castro-Ginard}, A. and {Chaoul}, L. and {Charlot}, P. and {Chemin}, L. and {Chiaramida}, V. and {Chiavassa}, A. and {Chornay}, N. and {Comoretto}, G. and {Contursi}, G. and {Cooper}, W.~J. and {Cornez}, T. and {Cowell}, S. and {Crifo}, F. and {Cropper}, M. and {Crosta}, M. and {Crowley}, C. and {Dafonte}, C. and {Dapergolas}, A. and {David}, M. and {David}, P. and {de Laverny}, P. and {De Luise}, F. and {De March}, R.},
        title = "{Gaia Data Release 3. Summary of the content and survey properties}",
      journal = {\aap},
         year = 2023,
        month = jun,
       volume = {674},
          eid = {A1},
        pages = {A1},
          doi = {10.1051/0004-6361/202243940},
archivePrefix = {arXiv},
       eprint = {2208.00211},
 primaryClass = {astro-ph.GA},
       adsurl = {https://ui.adsabs.harvard.edu/abs/2023A&A...674A...1G}
}

@ARTICLE{gallart2019,
       author = {{Gallart}, Carme and {Bernard}, Edouard J. and {Brook}, Chris B. and
         {Ruiz-Lara}, Tom{\'a}s and {Cassisi}, Santi and {Hill}, Vanessa and
         {Monelli}, Matteo},
        title = "{Uncovering the birth of the Milky Way through accurate stellar ages with Gaia}",
      journal = {\natas},
         year = 2019,
        month = jul,
       volume = {3},
        pages = {932-939},
          doi = {10.1038/s41550-019-0829-5},
archivePrefix = {arXiv},
       eprint = {1901.02900},
 primaryClass = {astro-ph.GA},
       adsurl = {https://ui.adsabs.harvard.edu/abs/2019NatAs...3..932G}
}

@ARTICLE{Bonaca2020,
       author = {{Bonaca}, Ana and {Conroy}, Charlie and {Cargile}, Phillip A. and {Naidu}, Rohan P. and {Johnson}, Benjamin D. and {Zaritsky}, Dennis and {Ting}, Yuan-Sen and {Caldwell}, Nelson and {Han}, Jiwon Jesse and {van Dokkum}, Pieter},
        title = "{Timing the Early Assembly of the Milky Way with the H3 Survey}",
      journal = {\apjl},
         year = 2020,
        month = jul,
       volume = {897},
       number = {1},
          eid = {L18},
        pages = {L18},
          doi = {10.3847/2041-8213/ab9caa},
archivePrefix = {arXiv},
       eprint = {2004.11384},
 primaryClass = {astro-ph.GA},
       adsurl = {https://ui.adsabs.harvard.edu/abs/2020ApJ...897L..18B}
}

@ARTICLE{DeasonBelokurov2024rev,
       author = {{Deason}, Alis J. and {Belokurov}, Vasily},
        title = "{Galactic Archaeology with Gaia}",
      journal = {\nar},
         year = 2024,
        month = dec,
       volume = {99},
          eid = {101706},
        pages = {101706},
          doi = {10.1016/j.newar.2024.101706},
archivePrefix = {arXiv},
       eprint = {2402.12443},
 primaryClass = {astro-ph.GA},
       adsurl = {https://ui.adsabs.harvard.edu/abs/2024NewAR..9901706D}
}

@ARTICLE{Haywood2018,
       author = {{Haywood}, M. and {Di Matteo}, P. and {Lehnert}, M.~D. and {Snaith}, O. and {Khoperskov}, S. and {G{\'o}mez}, A.},
        title = "{In Disguise or Out of Reach: First Clues about In Situ and Accreted Stars in the Stellar Halo of the Milky Way from Gaia DR2}",
      journal = {\apj},
         year = 2018,
        month = aug,
       volume = {863},
       number = {2},
          eid = {113},
        pages = {113},
          doi = {10.3847/1538-4357/aad235},
archivePrefix = {arXiv},
       eprint = {1805.02617},
 primaryClass = {astro-ph.GA},
       adsurl = {https://ui.adsabs.harvard.edu/abs/2018ApJ...863..113H}
}

@ARTICLE{Lane2023_GSEmass,
       author = {{Lane}, James M.~M. and {Bovy}, Jo and {Mackereth}, J. Ted},
        title = "{The stellar mass of the Gaia-Sausage/Enceladus accretion remnant}",
      journal = {\mnras},
         year = 2023,
        month = nov,
       volume = {526},
       number = {1},
        pages = {1209-1234},
          doi = {10.1093/mnras/stad2834},
archivePrefix = {arXiv},
       eprint = {2306.03084},
 primaryClass = {astro-ph.GA},
       adsurl = {https://ui.adsabs.harvard.edu/abs/2023MNRAS.526.1209L}
}

@article{Partmann_2025,
    author = {Partmann, Christian and Naab, Thorsten and Lahén, Natalia and Rantala, Antti and Hirschmann, Michaela and Hislop, Jessica M and Petersson, Jonathan and Johansson, Peter H},
    title = {The importance of nuclear star clusters for massive black hole growth and nuclear star formation in simulated low-mass galaxies},
    journal = {Monthly Notices of the Royal Astronomical Society},
    volume = {537},
    number = {2},
    pages = {956-977},
    year = {2025},
    month = {02},
    issn = {0035-8711},
    doi = {10.1093/mnras/staf002},
    url = {https://doi.org/10.1093/mnras/staf002},
    eprint = {https://academic.oup.com/mnras/article-pdf/537/2/956/61592299/staf002.pdf},
}

@ARTICLE{Chandrasekhar1943dynfriction,
       author = {{Chandrasekhar}, S.},
        title = "{Dynamical Friction. I. General Considerations: the Coefficient of Dynamical Friction.}",
      journal = {\apj},
         year = 1943,
        month = mar,
       volume = {97},
        pages = {255},
          doi = {10.1086/144517},
       adsurl = {https://ui.adsabs.harvard.edu/abs/1943ApJ....97..255C}
}
\bibliographystyle{aasjournal}



\end{document}